\documentclass[lettersize,journal]{IEEEtran}

\usepackage[english]{babel}
\usepackage{amsmath}
\usepackage{amssymb}
\usepackage{graphicx}
\usepackage{booktabs}
\usepackage{multirow}
\usepackage{xr-hyper}
\usepackage[colorlinks=true, allcolors=blue]{hyperref}
\usepackage[table,dvipsnames]{xcolor}
\usepackage{enumitem}
\usepackage{algorithm}
\usepackage{braket}
\usepackage{algpseudocode}
\usepackage[compatibility=false]{caption}
\usepackage{subcaption}
\usepackage{comment}
\usepackage{makecell}
\usepackage{xspace}
\usepackage{balance}
\usepackage{microtype}
\usepackage{listings}
\usepackage{url}
\usepackage[numbers]{natbib}
\usepackage{standalone}
\usepackage{orcidlink}

\usepackage{tabularx}
\usepackage{array}
\newcolumntype{Y}{>{\raggedright\arraybackslash}X}

\usepackage{tikz}
\usetikzlibrary{positioning, arrows.meta, fit,
                 backgrounds, calc, decorations.pathreplacing, automata}

\newcommand{\sys}{\textsc{HPC-vQPU}\xspace}
\newcommand{\sysserver}{\textsc{vQPU-Server}\xspace}
\newcommand{\sysagent}{\textsc{vQPU-Agent}\xspace}

\newcommand{\Queued}{\textsc{Queued}\xspace}
\newcommand{\Running}{\textsc{Running}\xspace}
\newcommand{\Completed}{\textsc{Completed}\xspace}
\newcommand{\Failed}{\textsc{Failed}\xspace}
\newcommand{\Cancelled}{\textsc{Cancelled}\xspace}

\definecolor{lightgray}{gray}{0.92}

\title{\sys: A Service-Export Architecture for Virtual QPUs on Batch-Scheduled HPC Systems}

\author{%
\IEEEauthorblockN{
Shusen Liu\orcidlink{0000-0002-0794-1132}\IEEEauthorrefmark{1}\IEEEauthorrefmark{2},
Pascal Jahan Elahi\orcidlink{0000-0002-6154-7224}\IEEEauthorrefmark{1}\IEEEauthorrefmark{2},
Ugo Varetto\orcidlink{0000-0002-7696-0345}\IEEEauthorrefmark{1}\IEEEauthorrefmark{2}
}

\IEEEauthorblockA{
\IEEEauthorrefmark{1}Pawsey Supercomputing Research Centre, Australia
}

\IEEEauthorblockA{
\IEEEauthorrefmark{2}The University of Western Australia, Australia
}

\IEEEauthorblockA{
Email: Shusen.liu@pawsey.org.au \& pascal.elahi@pawsey.org.au
}
}

\begin{document}

\maketitle

\begin{abstract}

Device-aware quantum simulation increasingly requires HPC-scale accelerators, yet secure supercomputers expose batch-scheduled execution environments rather than the interactive, backend-oriented interfaces expected by quantum software. The key obstacle is not only remote job submission: an HPC-hosted virtual QPU must preserve topology, native-gate, and calibration semantics across queue delay, scheduler allocation, compute-node isolation, and partial execution-side failures, without opening inbound paths into the cluster.

We present \sys, a service-export architecture for virtual QPUs on batch-scheduled HPC systems. \sys separates a cloud-facing control plane, which owns device identity, task lifecycle, snapshot binding, and event projection, from an HPC-resident execution plane, which claims work and realises it through scheduler-backed GPU jobs. Coordination is exclusively outbound and agent initiated. The central abstraction is a topology- and calibration-aware device snapshot bound atomically at claim time and carried into execution as an immutable contract, making each scheduled job hermetic while preserving fresh device semantics.

We implement \sys at the Pawsey Supercomputing Research Centre using Setonix GPUs, Qiskit-Aer/cuQuantum, and IBM Fez calibration data. Production experiments show that service overhead is bounded and additive, while workload scaling remains confined to the simulator; calibration-bearing snapshots produce measurable output shifts; claim-time binding prevents stale execution after pre-claim device mutation; concurrent agents complete 50/50 tasks exactly once; and explicit recovery restores stale running tasks after agent failure. These results show that secure, scheduler-mediated HPC infrastructure can export device-faithful quantum simulation as an interactive virtual-QPU service.

\end{abstract}



\section{Introduction}
\label{sec:intro}

Quantum circuit simulation remains indispensable in the NISQ era~\cite{preskill2018quantum}. Physical QPUs (Quantum Processor Units) continue to improve, but they remain scarce, noisy, calibration-sensitive, connectivity-limited, and difficult to access at the cadence required by systems research. Classical simulation therefore remains a verification reference, a compiler and routing testbed, a noise-model laboratory, and a means of studying devices that do not yet physically exist~\cite{Paraskevopoulos2025, Haener2016}. In these workflows, the target is not merely an ideal backend. It is a device-aware execution model characterised by topology, native operations, calibration parameters, and hardware-specific constraints.

The demand for such simulation is increasing while managed cloud simulation is becoming more bounded. Commercial quantum platforms provide useful device-oriented APIs, but broad, open-ended simulation is expensive for providers and often subject to scale, functionality, or cost limits~\cite{IBMQuantum2024, AWS2026}. Research workloads such as compiler campaigns, routing experiments, calibration-sensitivity sweeps, noise-model analysis, and future-QPU prototyping require sustained accelerator access, large memory capacity, and orchestration across many jobs~\cite{babuji2019parsl, rocklin2015dask, deelman2015pegasus, wilde2011swift}. Modern supercomputers, and simulator stacks such as cuQuantum and Qiskit-Aer, are a natural substrate for this demand~\cite{rajaraman2023frontier, allen2025aurora, bayraktar2023cuquantum, steiger2018projectq}.

Yet computational suitability does not imply service-level usability. Quantum developers expect a device-oriented workflow: submit a circuit, name a backend, retrieve a structured result, and iterate programmatically~\cite{ibm_quantum_learning}. HPC systems expose a scheduler-oriented workflow: authenticate to a login node, prepare site-specific environments, write a batch script, submit to a queue, wait for allocation, inspect logs, and reconcile scheduler state with application intent. The mismatch is semantic as well as ergonomic. Quantum users reason about circuits, shots, topology, and device state; HPC schedulers reason about jobs, partitions, nodes, queues, and resource grants.

A simple web wrapper around an HPC simulator fails at this boundary. Production HPC facilities commonly enforce asymmetric connectivity: login nodes may initiate outbound communication but are not intended to host public services, and compute nodes are scheduler-allocated, internally addressed, and externally unreachable~\cite{ARCHER2Service2026, NERSCC2026, ORLCF2026, PSRC2026}. Opening inbound paths into a shared supercomputer expands the attack surface and conflicts with multi-tenant site policy~\cite{thain2005distributed, turilli2018comprehensive}. We call this the \emph{connectivity inversion problem}: the execution side can safely contact the service side, but a correct service cannot rely on initiating communication into the cluster.

A second challenge is semantic. A \emph{virtual QPU} is a software-defined endpoint that exposes topology, native gate set, calibration-derived noise parameters, and target-specific constraints while realising execution through a simulator~\cite{Zheng2026}. This abstraction is what makes simulation useful for compiler development, routing evaluation, noise studies, and future-device exploration. Exporting it through an HPC scheduler is non-trivial because the device meaning of a task must survive admission, queue delay, claim, node allocation, execution, and result collection.

A third challenge is operational. Batch-mediated execution expands the failure surface well beyond ordinary application errors: a task may be invalid, but a job may also never start, a node may fail mid-run, the software environment may drift, a result artifact may be malformed, or an execution agent may disappear while jobs remain in flight. A viable service must therefore reconcile cloud-style interactive semantics, HPC security and scheduling constraints, and end-to-end device fidelity.

We present \sys, a service-export architecture for virtual QPUs on batch-scheduled HPC systems. Its central thesis is that HPC-hosted device-aware simulation can be exported as an interactive service without weakening HPC security boundaries, provided that service authority is separated from scheduler-governed execution. A cloud-facing \emph{control plane} manages the API, task lifecycle, device state, snapshot binding, and event projection. An HPC-resident \emph{execution plane} manages claiming, scheduler interaction, runtime preparation, and compute-node execution through an unprivileged login-node agent. The two planes communicate exclusively through outbound, agent-initiated messages.

The key execution object is a topology- and calibration-aware \emph{device snapshot}. \sys binds the snapshot at claim time, when task ownership is committed, and carries it into the scheduled job as a self-contained execution contract. This point is late enough to reflect device updates that occur after submission, early enough to keep compute-node execution hermetic, and general enough to support both current-device emulation and hypothetical future-device topologies.

\sys is therefore neither a thin REST adapter over Slurm~\cite{yoo2003slurm,SchedMD2026} or PBS~\cite{OpenPBSProject2026}, nor a generic science gateway layered on top of a simulator. It is a service-execution architecture in which device identity, calibration freshness, and lifecycle authority are first-class objects of the cross-boundary contract rather than annotations attached to a job.

The main contributions are:

\begin{enumerate}
\item \textbf{A service-export architecture for secure, batch-scheduled HPC systems.} We introduce a control-plane/execution-plane decomposition that resolves connectivity inversion through outbound-only, agent-initiated coordination while preserving the site's security posture and hiding scheduler details from clients.

\item \textbf{A device snapshot abstraction as a portable execution contract.} We define a graph-structured, topology- and calibration-aware representation of a virtual device that is bound to a task at claim time and carried into scheduled execution as a self-contained input.

\item \textbf{A scheduler-aware task lifecycle with bounded recovery.} We define a five-state task automaton and an asymmetric claim/report/heartbeat protocol that reconcile interactive service semantics with batch-mediated execution and partially observable failures.

\item \textbf{A reference implementation and production validation.} We implement \sys in the Pawsey environment using Qiskit-Aer with cuQuantum and evaluate it through regression tests and end-to-end experiments that quantify overhead, demonstrate calibration fidelity with IBM Fez data, confirm claim-time binding, and verify task integrity and recovery under production scheduling.
\end{enumerate}

The remainder of the paper is organised as follows. Section~\ref{sec:background} formalises the problem and constraints. Section~\ref{sec:overview} presents the architecture and task flow. Section~\ref{sec:design} develops the nine design invariants. Section~\ref{sec:impl} maps those invariants to the implementation. Section~\ref{sec:eval} evaluates the production deployment. Sections~\ref{sec:related} and~\ref{sec:discussion} position the work and discuss scope.


\section{Background and Problem Formulation}
\label{sec:background}

\subsection{Device-Aware Simulation as an HPC Service}

Quantum simulation has become a serious HPC workload. State-vector simulation of an $n$-qubit circuit requires storing $2^n$ complex amplitudes, while density-matrix models, stochastic noise, repeated shots, calibration sweeps, and benchmark campaigns further increase computational and memory demand. Simulator ecosystems reflect this shift: cuQuantum, Qiskit-Aer, and SV-Sim all target GPU or multi-node HPC execution rather than only local prototyping~\cite{bayraktar2023cuquantum,QAD2026,JavadiAbhari2024,Li2021,Lykov2021}. Appendix~\ref{supp-app:background-motivation} preserves the fuller simulator and access-model context.

Scale alone is not the defining requirement. Current quantum systems research increasingly depends on device structure. Compilation quality depends on coupling maps and native gate sets~\cite{Li2019,JavadiAbhari2024}; routing quality depends on geometry and directionality~\cite{Li2019,Murali2019}; result quality depends on calibration-sensitive noise~\cite{Kurniawan2024}; and practical future-device studies vary topology, native operations, and error parameters before hardware exists~\cite{preskill2018quantum,Haener2017,Paraskevopoulos2025}. A useful simulator is therefore not simply a place to run a circuit. It is a stand-in for a concrete or hypothetical processor.

This shift changes the service abstraction. Earlier workflows could often treat simulation as an ideal backend selected for convenience. Modern workflows cannot. Error mitigation, routing, layout selection, and architecture exploration depend on decoherence, gate infidelity, measurement bias, and connectivity constraints~\cite{Cai2023,Quek2024,Zhang2025}. Access to real QPU time is limited and costly~\cite{AWS2025,IBMQuantum2026a,Ma2025}, and future architectures are expensive to fabricate~\cite{Paraskevopoulos2025}. Many compiler, mapping, and runtime decisions must therefore be evaluated before scarce hardware access is consumed or before a proposed chip exists.

This motivates a \emph{virtual device service}: a service that exposes stable device identity, machine-readable topology, native operations, calibration-aware execution semantics, circuit submission, lifecycle state, and structured results. Such a service should let users move naturally among three modes of work: emulating an existing physical processor, studying a particular calibration snapshot, and exploring hypothetical architectures by varying topology, gate set, or error parameters. The simulator is then not an ideal backend hidden behind an API; it is a programmable virtual QPU whose device semantics are first-class objects that can be queried, selected, versioned, and used for systematic experimentation.

The variety of possible targets makes this a systems problem rather than a convenience feature. Even within a single provider, multiple device generations can coexist with different connectivity, native operations, calibration refresh rates, and disabled-qubit patterns. Across future-device studies, the target may not correspond to any vendor backend at all. A user may therefore want to ask not only ``can this circuit run?'' but ``which current or proposed device makes this circuit meaningful?'' A device-agnostic simulator cannot answer that question because it removes the very structure being evaluated.

\subsection{Why Existing Access Models Are Insufficient}

Traditional HPC and commercial cloud access optimise for different invariants. HPC systems provide the tightly coupled accelerator substrate needed for large simulation, but expose it through accounts, filesystems, batch scripts, modules, queues, and allocation policies rather than through device-level service contracts~\cite{yoo2003slurm,PSRC2023}. This is appropriate for conventional batch science, but poorly aligned with external tools that want to address a named backend, submit circuits programmatically, and observe lifecycle state through a structured API.

Cloud platforms invert the problem. Their APIs provide the device-oriented boundary that quantum users expect, but large communication-heavy or bursty simulation reintroduces explicit concerns about placement, capacity, storage, and cost~\cite{Armbrust2009,AWS2026a,AWS2026b}. Large-scale, device-aware quantum simulation therefore requires a different access model: managed device semantics at the boundary, with execution realised on scheduler-managed, accelerator-rich HPC infrastructure.

This distinction matters because the service object determines what must be preserved. If the service object is a generic job, then success means that a script was submitted and some output was returned. If the service object is a virtual device, then success means that the circuit was admitted against a specific device description, executed under a well-defined topology and calibration state, and returned through a lifecycle whose meaning is independent of scheduler transients. The latter is the problem addressed in this paper.

This is why a generic science gateway or job portal is insufficient by itself. Such systems can simplify remote execution, but they do not normally make calibration freshness, topology admissibility, native-gate constraints, and device identity part of the execution contract. For \sys, these properties are not annotations around a job; they determine whether the result is scientifically interpretable as execution on the requested virtual QPU.

\subsection{Connectivity Inversion}

At first glance, one might place an API outside the cluster and have it dispatch work to the execution side. In the target HPC setting, this control-side-reaches-worker model is structurally unavailable. Login nodes may initiate outbound communication but are not public services. Compute nodes are scheduler-allocated, internally addressed, transient, and typically unreachable from external networks~\cite{ARCHER2Service2026,NERSCC2026,ORLCF2026,PSRC2026}. A broker or service inside the cluster would itself require inbound reachability or privileged network configuration.

Opening such paths would weaken the site's security and audit model. Operators would need to reason not only about which job ran, but which external request triggered it, under which service identity, and how that maps to scheduler accounting. This conflicts with the multi-tenant nature of HPC systems, where identity, allocation, and compliance are mediated through established site mechanisms~\cite{thain2005distributed,turilli2018comprehensive}. We call this the \emph{connectivity inversion problem}: the execution side may safely initiate communication outward, but a correct design must not rely on initiating communication inward.

The inversion is compounded by the natural multiplicity of the service relationship. The service is not coordinating one stable worker. It may coordinate several login-node agents, many scheduler jobs, and transient compute-node runners created and destroyed by the scheduler. A push-based design would need stable addressing, secure invocation, auditability, and lifecycle coordination across this changing population. In the target setting, the safer direction is to make execution-side actors pull work outward and let the control plane remain a public service endpoint rather than a cluster ingress mechanism.

\subsection{Device State and Admissibility}

A task in a virtual-QPU service must preserve the semantics of the target device, not merely the syntax of the circuit. We model a virtual device $\mathcal{D}$ as
\begin{equation}
\mathcal{D} = (Q, G, \mathcal{G}_{\mathrm{native}}, \Theta_Q, \Theta_E),
\end{equation}
where $Q$ is the qubit set, $G=(Q,E)$ is the possibly directed coupling graph, $\mathcal{G}_{\mathrm{native}}$ is the native gate set, and $\Theta_Q,\Theta_E$ are per-qubit and per-edge calibration fields. At runtime, \sys uses a time-indexed snapshot
\begin{align}
\Delta(\mathcal{D}, t) = \Bigl(
&G, \mathcal{G}_{\mathrm{native}},\notag \\
&\{(T_1(q_i),T_2(q_i),\epsilon_{\mathrm{1q}}(q_i),r(q_i))\}_{q_i \in Q}, \notag\\
&\{(\epsilon_{ij},g_{ij})\}_{(q_i,q_j)\in E}
\Bigr).
\label{eq:snapshot}
\end{align}
Here $T_1,T_2$ capture coherence, $\epsilon_{\mathrm{1q}}$ and $\epsilon_{ij}$ capture one- and two-qubit gate error, $r$ captures readout error, and $g_{ij}$ records the native two-qubit operation on an edge. Appendix~\ref{supp-app:device-model-details} gives the full field interpretation and the rationale for keeping the representation graph-structured rather than vendor-specific.

This representation is deliberately more than metadata. If the service transmitted only a circuit and device name, then a scheduled job would have to recover device meaning later through a live lookup. Under queue delay and compute-node isolation, that dependency is fragile and semantically ambiguous: the live device state may have changed after submission, or the compute node may not be able to reach the server at all. Conversely, if the service ignores device state and runs only an ideal simulator, it loses the fidelity required by the workflows motivating this paper. The snapshot is the minimal object that makes device meaning explicit, portable, and stable enough to survive scheduler-mediated execution.

Given a task $\tau=(c,\ell,n_s,d)$, where $c$ is a circuit, $\ell$ its dialect, $n_s$ the shot count, and $d$ the target device, the service checks whether $c$ is device-admissible with respect to the advertised device description. A circuit is admissible only if every referenced qubit is operational, every operation belongs to $\mathcal{G}_{\mathrm{native}}$, and every multi-qubit operation follows an available coupling in $G$. This admission-time check rejects impossible requests before scheduler resources are committed; the immutable execution snapshot is bound later at claim time.

This separation between admission and execution is subtle but central. Admission establishes that a request is meaningful for the advertised virtual device; it does not choose the final calibration state. Claim establishes ownership and binds the concrete snapshot that will be executed. The design therefore avoids two weaker alternatives. Binding at submission would make long-queued tasks stale by construction when calibration changes before any execution resource is available. Binding inside the compute-node runner would require a live lookup from an isolated job and would make task meaning depend on network reachability and mutable state. The admissibility relation is intentionally stable enough to reject structurally impossible circuits early, while the calibration-bearing snapshot remains fresh until the moment the execution side actually accepts responsibility.

\subsection{Expanded Failure Surface}

Crossing the scheduler boundary expands the failure model. Tasks may fail because a circuit is invalid, but also because a scheduler job never starts, a node fails mid-run, a software environment drifts, a result artifact is missing or malformed, or an agent disappears while work remains in flight. Many of these failures occur in systems only partially observable to the control plane. A service-export architecture must therefore preserve a client-visible lifecycle, exclusive task ownership, liveness observability, explicit recovery, and device semantics without live server contact during compute-node execution.

\subsection{Problem Statement and Constraints}
\label{sec:problem_statement}

The system receives a stream of device-aware tasks
\[
\mathcal{W}=\langle \tau_1,\tau_2,\dots,\tau_k,\dots\rangle
\]
from users, tools, or automated workflows. It must realise a mapping $\mathcal{S}:\mathcal{W}\rightarrow\mathcal{R}$ that produces structured outcomes $\rho_i$ associated with the correct task, preserves the intended device semantics, and exposes a stable service interface independent of scheduler details. The problem is therefore not how to run one simulation on HPC, but how to export a concurrent virtual-device service backed by scheduler-mediated infrastructure.

The objective has two simultaneous sides. Externally, $\mathcal{S}$ must behave like a device-oriented service: tasks are submitted against named targets and results are interpreted against those targets. Internally, execution must remain faithful to HPC operation: jobs are scheduler allocated, compute nodes are not public services, and site-specific launch details remain below the service boundary. The architecture must therefore preserve device semantics while deliberately avoiding direct service control over the execution substrate.

The mapping is constrained by ownership and time. Let $owner(\tau_i,t)$ denote the execution-side agent, if any, that is authorised to realise $\tau_i$ at time $t$. Let $state(\tau_i,t)$ be the service-visible lifecycle state. A valid execution history must have a single linearisation point at which $state(\tau_i,t)$ changes from \Queued\ to \Running, $owner(\tau_i,t)$ becomes non-null, and $\delta_i=\Delta(\mathcal{D}_{d_i},t)$ is chosen. After this point, later mutations of $\mathcal{D}_{d_i}$ may affect future claims, but must not rewrite $\delta_i$ for the already claimed task. Similarly, terminal publication must be absorbing: once $\rho_i$ is committed as completed or failed, later reports for the same task can be recorded as evidence or rejected, but cannot create a second authoritative outcome. These requirements turn the service from a convenient submission wrapper into a lifecycle authority.

The system must satisfy six constraints:

\textbf{C1: No reliance on inbound control-plane connectivity.} The service must not initiate communication toward execution-side agents or compute resources; all cross-boundary coordination must be agent initiated.

\textbf{C2: Scheduler-mediated execution.} Every task must execute under a scheduler-allocated job, not through unmanaged compute-side daemons.

\textbf{C3: Device-semantic preservation.} For $\tau_i=(c_i,\ell_i,s_i,d_i)$, let $\delta_i=\Delta(\mathcal{D}_{d_i},t_{\mathrm{claim}}(\tau_i))$. The result must satisfy $\rho_i=\mathcal{E}(\tau_i,\delta_i)$ and remain invariant under later changes to $\mathcal{D}_{d_i}$.

\textbf{C4: Concurrency correctness.} A claimed task has at most one owner; concurrent agents must not duplicate execution, lose ownership, or create ambiguous result assignment.

\textbf{C5: Stable client-facing abstraction.} Users should interact with devices, tasks, and results, not scheduler job IDs, queue policies, module stacks, or site-specific procedures.

\textbf{C6: Operational observability and recovery.} Progress and failure must remain observable and recoverable at the service level despite scheduler delay, agent failure, and partial execution-side visibility.

The next section presents the architecture that satisfies C1--C6; Section~\ref{sec:design} states the invariants that make the satisfaction precise.

\section{System Overview}
\label{sec:overview}

\sys is organised around a boundary principle: service semantics and scheduler-mediated execution must remain in separate operational domains. The cloud-facing side reasons about devices, tasks, lifecycle state, snapshots, and events. The HPC-facing side reasons about local files, scheduler submission, run directories, and compute-node execution. The two sides share no process state, storage engine, runtime dependency, or cluster-specific assumption. They coordinate only through versioned, outbound, agent-initiated messages.

\begin{figure*}[!htp]
    \centering
    \includegraphics[width=0.8\textwidth]{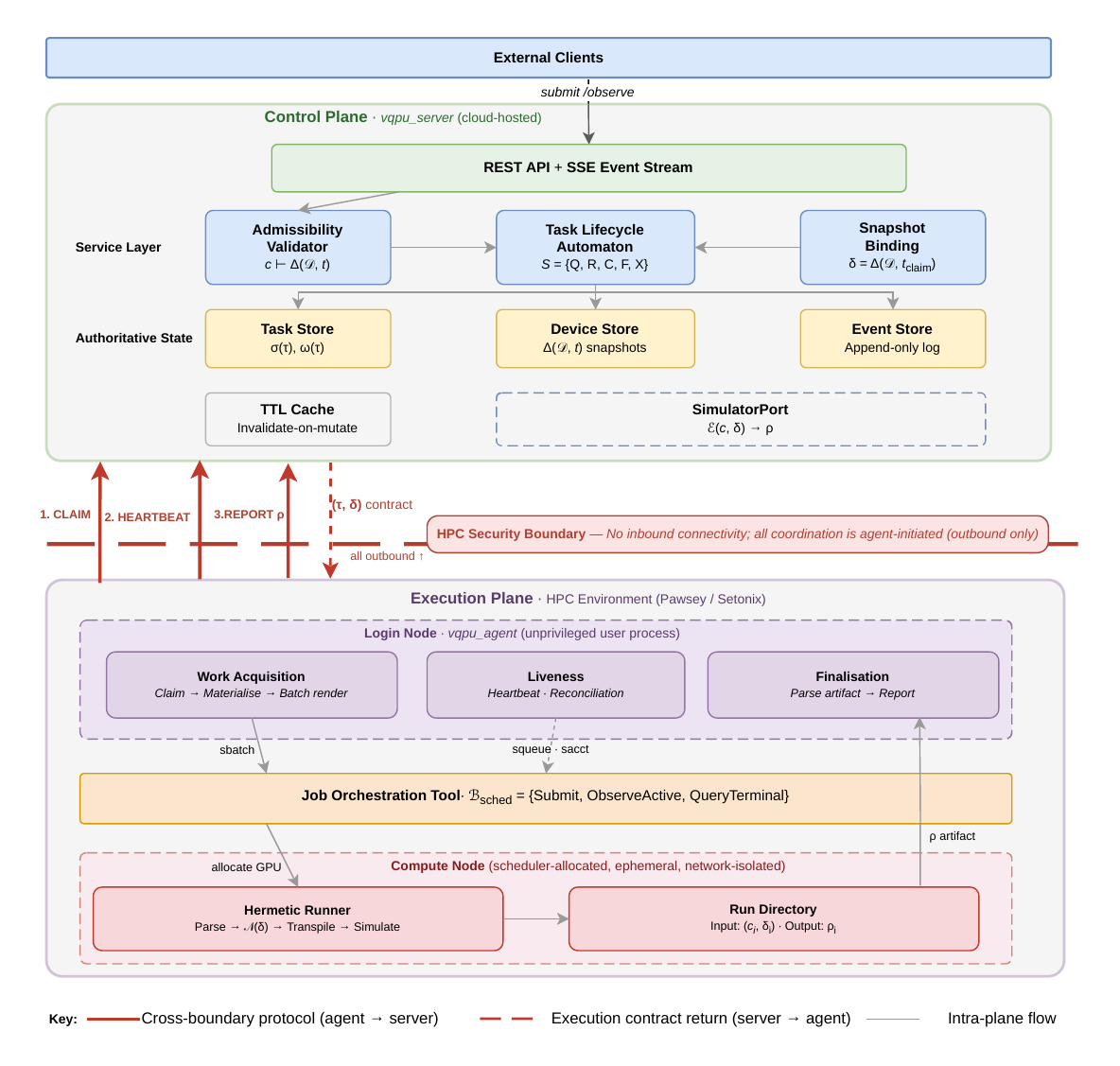}
    \caption{
    \sys architecture.
    The cloud-hosted control plane owns admissibility validation, task lifecycle state, claim-time snapshot binding, and authoritative task/device/event stores.
    The HPC-resident execution plane owns scheduler interaction and hermetic compute-node realisation through an unprivileged login-node agent.
    The planes are joined only by outbound, agent-initiated \textsc{Claim}, \textsc{Heartbeat}, and \textsc{Report} operations.
    At claim time, the server returns $(\tau_i,\delta_i)$, where $\delta_i=\Delta(\mathcal{D}_{d_i},t_{\mathrm{claim}}(\tau_i))$; the runner evaluates $\rho_i=\mathcal{E}(\tau_i,\delta_i)$ from its local payload.
    }
    \label{fig:arch}
\end{figure*}

This decomposition resolves connectivity inversion without weakening the HPC boundary. Instead of placing a reachable execution service inside the cluster, \sys lets the cluster export simulation capacity outward through an unprivileged login-node agent. The result is the two-plane architecture in Figure~\ref{fig:arch}: a \emph{control plane} that owns service authority and an \emph{execution plane} that owns scheduler-governed realisation.

\subsection{Control Plane}
\label{sec:overview-control}

The control plane is implemented by \sysserver, a cloud-accessible service that defines the system's external abstraction. Its role is not to run jobs directly, but to preserve a stable device-oriented service contract over delayed and partially observable scheduler-mediated execution.

First, the control plane performs synchronous acceptance. A client submits $\tau_i=(c_i,\ell_i,s_i,d_i)$ through \texttt{POST /tasks}; the server checks device admissibility, allocates a durable identifier, persists the task in \Queued, and returns a structured task record. This decouples client interaction from immediate scheduler availability: the service can accept work even when no agent is presently online or no GPU allocation is available.

Second, it is authoritative for task and device state. For tasks, it enforces lifecycle transitions, ownership, terminal-state absorption, and result publication. For devices, it stores the current set of virtual QPU snapshots, including topology, native gate set, calibration-derived parameters, and availability. The execution side may report runner identity, scheduler job identifiers, heartbeat timestamps, and result artifacts, but these are evidence submitted to the control plane; they do not replace the service state machine.

Third, it provides external observability. Every meaningful transition is emitted as a typed event and exposed through Server-Sent Events, allowing dashboards and automation to track progress without reaching into the HPC domain. The control plane is therefore deliberately scheduler-blind: it speaks in devices, tasks, lifecycle states, and results, not partitions, module stacks, or batch-script details.

\subsection{Execution Plane}
\label{sec:overview-exec}

The execution plane is implemented by \sysagent, an unprivileged process on an HPC login node. The agent establishes availability by participating in the protocol: it claims work and emits heartbeats bearing its runner identity. If it stops, absence becomes visible through heartbeat expiry rather than through any server-to-agent callback.

When local concurrency permits, the agent issues \textsc{Claim}. The server atomically transfers ownership and returns the bound execution payload $(\tau_i,\delta_i)$. The agent then materialises this payload into an isolated run directory, prepares site-local scheduler configuration, submits execution through a Dask-backed scheduler adapter, observes completion, reads the result artifact, and reports either \textsc{Report-Completed} or \textsc{Report-Failed}. Dask is used as an adapter over the site scheduler, not as a replacement for it; allocation remains scheduler-mediated.

The agent is intentionally reconstructible. It owns no durable service truth and does not manage devices, authentication, REST resources, or public ports. On restart, it rebuilds its working view from the control plane, scheduler evidence, and local run directories. This keeps the cluster-side component narrow and site-adaptable.

This role differs from a conventional long-running worker in three ways. First, the agent is an outbound participant in a protocol, not an addressable execution endpoint. Its disappearance is therefore detected through missed heartbeats and unfinished owned tasks rather than through a failed callback. Second, the scheduler remains the only mechanism that grants compute-node capacity. The agent may prepare a run directory and submit a job, but it does not reserve GPUs, bypass queue policy, or run a permanent compute-side daemon. Third, the compute-node runner is hermetic with respect to service semantics: it receives a local payload containing the circuit, shot count, dialect, and bound device snapshot, then writes a result artifact without needing to query the server. These constraints are what make the execution plane acceptable inside an ordinary HPC operational model.

The execution plane is also horizontally replicable. Multiple agents can poll the same control plane from the same or different login nodes, subject to local site policy and allocation limits. They do not coordinate through shared cluster state. Coordination occurs at the task claim operation, where the control plane serialises ownership. This keeps the concurrency primitive aligned with the semantic boundary: the system does not need a distributed lock around scheduler jobs, only a unique service-level owner for each device-aware task.

\subsection{Snapshot Contract and Task Flow}
\label{sec:overview-snapshot}

The device snapshot is the primary object transferred across the plane boundary. It gives the task a concrete scientific meaning: not merely ``run circuit $c$,'' but ``run circuit $c$ under this virtual device state.'' For calibration-sensitive workflows, two executions of the same circuit may legitimately differ because the coupling map, operational qubits, native gate on an edge, readout error, or two-qubit error changed. A snapshot pins those semantics to a concrete contract that can survive queue delay and compute-node isolation.


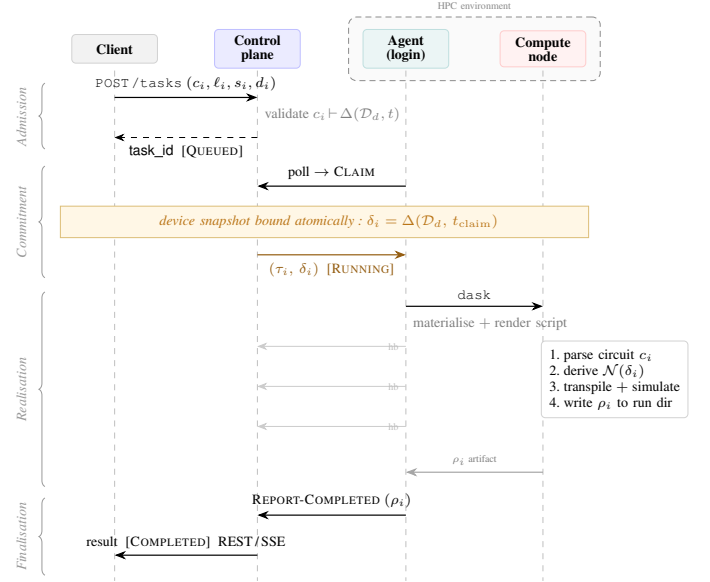
\begin{figure}[!htp]
\centering
\resizebox{0.50\textwidth}{!}{%
\begin{tikzpicture}[
  font=\scriptsize,
  >=Stealth,
  lifeline/.style={dashed, gray!50, very thin},
  msg/.style={->, semithick},
  rsp/.style={->, thin, dashed},
  lane/.style={font=\scriptsize\bfseries, minimum width=1.5cm,
               minimum height=0.5cm, rounded corners=2pt,
               inner sep=3pt, align=center, draw},
  phase/.style={font=\scriptsize\itshape, text=gray!80, rotate=90,
                anchor=south},
  phasebrace/.style={decorate,
    decoration={brace, amplitude=3pt, mirror, raise=2pt},
    gray!80, thin},
  hpcbox/.style={draw=gray!90, densely dashed, rounded corners=4pt,
                 fill=gray!4, inner sep=5pt},
]

\definecolor{amber}{RGB}{186,117,23}
\definecolor{snapfill}{RGB}{255,247,230}
\definecolor{snapborder}{RGB}{220,190,120}

\newcommand{\xC}{0cm}       
\newcommand{\xS}{2.5cm}     
\newcommand{\xA}{5.1cm}     
\newcommand{\xN}{7.5cm}     

\node[lane, fill=gray!10,   draw=gray!35]  (hCl) at (\xC,0) {Client};
\node[lane, fill=blue!8,    draw=blue!25]  (hSv) at (\xS,0) {Control\\[-1pt]plane};
\node[lane, fill=teal!8,    draw=teal!25]  (hAg) at (\xA,0) {Agent\\[-1pt](login)};
\node[lane, fill=red!5,     draw=red!20]   (hCn) at (\xN,0) {Compute\\[-1pt]node};

\begin{scope}[on background layer]
  \node[hpcbox, fit=(hAg)(hCn), inner ysep=6pt, inner xsep=7pt,
        label={[font=\tiny, gray!90, anchor=south]above:
               HPC environment}] (hpc) {};
\end{scope}

\foreach \x in {\xC,\xS,\xA,\xN}{
  \draw[lifeline] (\x,-0.38) -- (\x,-9.3);
}

\draw[phasebrace] (-1.15,-0.6) -- (-1.15,-1.75);
\node[phase] at (-1.42,-1.18) {Admission};

\draw[msg] (\xC,-0.85) --
  node[above, midway, font=\scriptsize]
    {\texttt{POST\,/tasks}\;$(c_i,\ell_i,s_i,d_i)$}
  (\xS,-0.85);

\node[font=\scriptsize, text=gray, align=center, anchor=west]
  at (\xS+0.15,-1.15) {validate\;$c_i\!\vdash\!\Delta(\mathcal{D}_d,t)$};

\draw[rsp] (\xS,-1.55) --
  node[below, midway, font=\scriptsize]
    {\textsf{task\_id}\;\;[\textsc{Queued}]}
  (\xC,-1.55);

\draw[phasebrace] (-1.15,-2.05) -- (-1.15,-4.0);
\node[phase] at (-1.42,-3.02) {Commitment};

\draw[msg] (\xA,-2.4) --
  node[above, midway, font=\scriptsize]
    {poll\;$\rightarrow$\;\textsc{Claim}}
  (\xS,-2.4);

\fill[snapfill]
  (-0.95,-2.75) rectangle (8.3,-3.3);
\draw[snapborder, thin]
  (-0.95,-2.75) rectangle (8.3,-3.3);
\node[font=\scriptsize\itshape, text=amber, anchor=center]
  at (3.75,-3.02)
  {device snapshot bound atomically\;:\;%
   $\delta_i = \Delta(\mathcal{D}_d,\,t_{\mathrm{claim}})$};

\draw[msg, amber!80!black, semithick] (\xS,-3.6) --
  node[below, midway, font=\scriptsize, text=amber!80!black]
    {$(\tau_i,\,\delta_i)$\;\;[\textsc{Running}]}
  (\xA,-3.6);

\draw[phasebrace] (-1.15,-4.25) -- (-1.15,-7.65);
\node[phase] at (-1.42,-5.95) {Realisation};

\draw[msg] (\xA,-4.5) --
  node[above, midway, font=\scriptsize]
    {\texttt{dask}}
  (\xN,-4.5);

\node[font=\scriptsize, text=gray, anchor=west]
  at (\xA+0.12,-4.8) {materialise\;$+$\;render script};

\node[draw=gray!35, thin, rounded corners=2pt,
      fill=white, align=left, font=\scriptsize,
      anchor=north west, inner sep=4pt,
      minimum width=2.0cm]
  (execbox) at (\xN-1.0,-5.1)
  {1.\;parse circuit $c_i$\\
   2.\;derive $\mathcal{N}(\delta_i)$\\
   3.\;transpile\;$+$\;simulate\\
   4.\;write $\rho_i$ to run dir};

\foreach \y/\lab in {-5.2/hb, -5.9/hb, -6.6/hb}{
  \draw[thin, gray!40, ->] (\xA,\y) -- (\xS,\y);
  \node[font=\tiny, text=gray!50, anchor=east] at (\xA-0.08,\y) {\lab};
}

\draw[msg, gray!70] (\xN,-7.4) --
  node[above, midway, font=\tiny, text=gray]
    {$\rho_i$\;artifact}
  (\xA,-7.4);

\draw[phasebrace] (-1.15,-7.85) -- (-1.15,-9.2);
\node[phase] at (-1.42,-8.52) {Finalisation};

\draw[msg] (\xA,-8.15) --
  node[above, midway, font=\scriptsize]
    {\textsc{Report-Completed}\;$(\rho_i)$}
  (\xS,-8.15);

\draw[msg] (\xS,-8.85) --
  node[above, midway, font=\scriptsize]
    {result\;\;[\textsc{Completed}]\;\;REST\,/\,SSE}
  (\xC,-8.85);

\end{tikzpicture}
}
\caption{End-to-end task flow through the four lifecycle phases.
  The client interacts only with the control plane; the agent
  mediates all scheduler and compute-node interaction from within
  the HPC environment.
  The shaded band marks the critical commitment boundary:
  the device snapshot
  $\delta_i = \Delta(\mathcal{D}_d, t_{\mathrm{claim}})$
  is bound atomically at \textsc{Claim} time and carried into
  the scheduled job as the execution contract
  $(\tau_i, \delta_i)$.
  Once committed, the compute-node runner evaluates
  $\rho_i = \mathcal{E}(c_i, \delta_i)$ hermetically
  from its local payload, with no further control-plane contact.
  Heartbeats~(hb) maintain bounded liveness
  during scheduler-mediated execution.}
\label{fig:sequence}
\end{figure}

Figure~\ref{fig:sequence} shows the full task sequence. It is useful to read the flow as four phases.

\textbf{Admission} is decoupled ingestion. The client submits a device-aware request; the control plane validates it against the currently advertised device description, assigns a durable identifier, and persists the task in \Queued. No HPC capacity is required at this point, which keeps the API responsive even when the scheduler is busy or no agent is currently available.

\textbf{Commitment} is the critical cross-domain handoff. An agent issues \textsc{Claim}; the server atomically moves the task from \Queued\ to \Running, assigns ownership, and binds the current snapshot $\delta_i$. After this point, the task is no longer a mutable queue element. It is an owned work unit with fully determined execution semantics.

\textbf{Realisation} occurs entirely inside the execution plane. The agent materialises $(\tau_i,\delta_i)$ into a local run directory and delegates execution to the scheduler adapter. The compute-node runner consumes the local payload, reconstructs topology and noise behaviour from the snapshot, and performs simulation without contacting the control plane. Scheduler delay may postpone execution, but it cannot alter the contract already bound at claim.

\textbf{Finalisation} moves the outcome back to the service domain. The agent reads the result artifact and reports either $\rho_i$ or a typed failure. The control plane commits the terminal transition, persists the outcome, and emits an event. This completes the round trip from device-oriented request to device-oriented result while preserving the HPC security boundary.

The flow uses only two cross-boundary handoffs carrying task semantics: the snapshot-bearing execution contract moves inward at claim time, and the structured result moves outward at completion.

\section{Design Invariants}
\label{sec:design}
The constraints formulated in Section~\ref{sec:background} and the two-plane decomposition introduced in Section~\ref{sec:overview} together define the design space. This section resolves that space into architectural invariants. Each subsection addresses a specific tension that arises when an interactive, device-aware service must be realised over inward-closed, scheduler-mediated infrastructure. In each case, the resolution is expressed not as a feature but as an invariant: a property the architecture must preserve regardless of scheduler state, agent availability, or device evolution. Nine such invariants (D1--D9) are established below. Together, they constitute the design contract that the implementation realises (Section~\ref{sec:impl}) and the evaluation validates (Section~\ref{sec:eval}).

\subsection{D1: Outbound-Only Connectivity Model}
\label{sec:design-connectivity}
C1 forbids any control-plane-initiated communication path from $\mathcal{S}$ to execution-side resources. A conventional distributed service would push tasks to reachable workers or route through a mutually reachable broker. Neither pattern survives in the target setting: compute nodes are scheduler-allocated, transiently addressed, and externally unreachable; a broker inside the cluster would itself require either inbound connectivity or privileged network configuration. \sys resolves this by reversing the direction of coordination. All cross-boundary communication is initiated outward by a login-node agent over standard outbound application protocols, such as HTTPS. The agent polls for work, claims tasks, and reports outcomes; the control plane never opens a connection into the cluster. This introduces bounded polling delay, but aligns with the network behaviour already permitted to unprivileged HPC users and requires no site-level security exceptions.

\textbf{Invariant D1.} Every cross-boundary interaction is initiated from within the HPC environment. \hfill\emph{(discharges C1)}

\subsection{D2: Scheduler-Aware Task Lifecycle}
\label{sec:design-statemachine}

\begin{figure}[t]
\centering
\resizebox{0.50\textwidth}{!}{%
\begin{tikzpicture}[
  >=Stealth,
  font=\small,
  every state/.style={
    minimum size=1.1cm,
    inner sep=2pt,
    font=\small\bfseries,
    rounded corners=3pt,
    draw,
    thick,
  },
  active/.style={
    fill=blue!6, draw=blue!40,
  },
  terminal/.style={
    fill=gray!10, draw=gray!50,
    double, double distance=1.2pt,
  },
  every edge/.style={draw, ->, semithick},
  op/.style={font=\scriptsize, fill=white, inner sep=1.5pt},
  recov/.style={densely dashed, gray!70},
  admin/.style={densely dotted, red!50!gray},
]

\node[state, active]   (Q) at (0, 0)    {\Queued};
\node[state, active]   (R) at (3.2, 0)  {\Running};
\node[state, terminal] (C) at (6.4, -2.6)  {\Completed};
\node[state, terminal] (F) at (3.2,-2.6){\Failed};
\node[state, terminal] (X) at (0, -2.6) {\Cancelled};



\draw (Q) edge[bend left=10]
  node[op, above] {\textsc{Claim}} (R);

\draw (R) edge[bend left=10]
  node[op, above] {\textsc{Report-Completed}} (C);

\draw (R) edge[]
  node[op, right] {\textsc{Report-Failed}} (F);


\draw (Q) edge[]
  node[op, left, pos=0.4] {\textsc{Cancel}} (X);

\draw (R) edge[bend left=12]
  node[op, above, pos=0.4] {\textsc{Cancel}} (X);


\draw[recov] (R) edge[bend left=20]
  node[op, below, font=\scriptsize\itshape] {\textsc{Requeue}} (Q);

\draw[admin] (Q) edge[bend left=10]
  node[op, below left=-10pt, font=\scriptsize\itshape, pos=0.39] {\textsc{Force-Fail}} (F);

\draw[admin] (R) edge[bend right=20]
  node[op, left=-4pt, font=\scriptsize\itshape, pos=0.65] {\textsc{Force-Fail}} (F);

\node[anchor=north west, font=\scriptsize, text=gray!70,
      align=left, inner sep=0pt]
  at (-0.8, -3.8)
  {%
   \tikz[baseline=-0.5ex]\draw[semithick,->](0,0)--(.5,0);~normal%
   \qquad
   \tikz[baseline=-0.5ex]\draw[recov,->](0,0)--(.5,0);~recovery%
   \qquad
   \tikz[baseline=-0.5ex]\draw[admin,->](0,0)--(.5,0);~administrative%
   \qquad
   \tikz[baseline=-0.5ex]\node[state,terminal,minimum size=5pt,
         inner sep=0pt,font=\relax]{};~terminal%
  };

\end{tikzpicture}
}
\caption{Task lifecycle automaton $S=\{\Queued, \Running, \Completed, \Failed, \Cancelled\}$.
  Solid arrows are agent-initiated protocol transitions;
  dashed arrows denote recovery (\textsc{Requeue});
  dotted arrows denote administrative intervention (\textsc{Force-Fail}).
  Double-bordered states are absorbing:
  once a task enters $\Completed$, $\Failed$, or $\Cancelled$,
  no further transition is permitted (Invariant~D2).}
\label{fig:statemachine}
\end{figure}
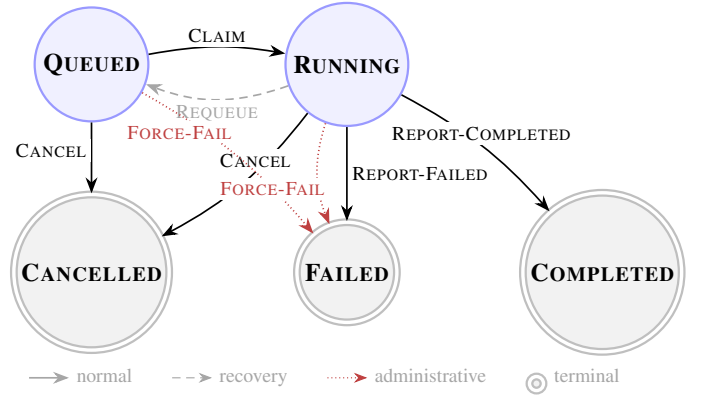

C5 requires that the client-facing abstraction remain device- and task-oriented rather than scheduler-oriented. Yet the execution path traverses a richer scheduler state machine whose states are site-specific, transient, and insufficient on their own: a task may be accepted by the service before any scheduler job exists, and may fail after the scheduler job terminates if result artifacts are missing or malformed. Exposing raw scheduler states would produce a brittle, deployment-dependent contract. \sys therefore interposes a deliberately compact service-level automaton
\begin{equation}
\begin{aligned}
S &= \{\Queued,\Running,\Completed,\\
   &\qquad \Failed,\Cancelled\},\\
F &= \{\Completed,\Failed,\Cancelled\}\subset S.
\end{aligned}
\label{eqn:task-lifecycle}
\end{equation}
whose transition relation is given in Figure~\ref{fig:statemachine}. The service owns this automaton independently of any intermediate scheduler states traversed beneath it. States in $F$ are absorbing: once a task reaches a terminal state, no further transition is permitted.



This automaton is a usability simplification and the mechanism by which client-visible semantics are decoupled from scheduler-specific behaviour. The scheduler remains an execution substrate; the five-state lifecycle remains the authoritative external contract.

\textbf{Invariant D2.} Terminal states are absorbing, and only terminal states constitute client-visible outcomes. \hfill\emph{(discharges C5)}

\subsection{D3: Exclusive Claim and Ownership Semantics}
\label{sec:design-claim}
C4 requires ownership to be unique whenever it is defined. We model ownership as a partial function $\omega : \mathcal{W} \rightharpoonup \mathrm{Agents}$, where $\omega(\tau_i)=a$ means that task $\tau_i$ is currently owned by agent $a$. When multiple agents poll concurrently, a naive fetch-then-update pattern can violate this requirement in practice: two agents may observe the same queued $\tau_i$ and attempt to execute it independently, wasting scarce GPU resources and producing ambiguous results. Task acquisition in \sys is therefore defined as an atomic \textsc{Claim} operation. \textsc{Claim} simultaneously selects a queued task, transitions it to \Running, and binds it to a specific owner identity, so that ownership is established at the same instant $\tau_i$ leaves \Queued. The architectural requirement is that ownership binding and state transition be indivisible; the concrete store mechanism that realises this requirement is an implementation question deferred to Section~\ref{sec:impl-claim} and Appendix~\ref{supp-app:design-claim-store}.

\textbf{Invariant D3.} Ownership binding is unique and established atomically with the \Queued\ $\rightarrow$ \Running\ transition. For every $\tau_i \in \mathcal{W}$, $\omega(\tau_i)$ is either undefined or equal to exactly one agent identity. \hfill\emph{(discharges C4)}

This invariant is foundational: without exclusive ownership, heartbeat interpretation, result attribution, cancellation, and recovery all become ambiguous.

\subsection{D4: Asymmetric Coordination Protocol}
\label{sec:design-protocol}

D1 prohibits the control plane from pushing state inward; C6 requires that execution-side liveness remain observable and that recovery be bounded. Together, they demand an agent-initiated protocol rich enough to distinguish operationally distinct situations: a task that has been claimed but not yet submitted to the scheduler, a task whose scheduler job is running, and a task whose agent has silently disappeared, without any server-to-agent callback. \sys therefore defines a narrow asymmetric protocol (Table~\ref{tab:protocol}) comprising five agent-initiated operations (\textsc{Claim}, \textsc{Report-Running}, \textsc{Report-Completed}, \textsc{Report-Failed}, \textsc{Heartbeat}) and two administrator-initiated recovery operations (\textsc{Requeue}, \textsc{Force-Fail}).

\begin{table}[ht]
\centering
\caption{Agent--server protocol in \sys.}
\label{tab:protocol}
\small
\setlength{\tabcolsep}{4pt}
\begin{tabular}{@{}l l p{0.56\columnwidth}@{}}
\toprule
Operation & Initiator & Purpose \\
\midrule
\textsc{Claim} & Agent & Atomically acquire next queued task \\
\makecell[l]{\textsc{Report}\\\textsc{-Running}} & Agent & Bind scheduler metadata to owned task \\
\makecell[l]{\textsc{Report}\\\textsc{-Completed}} & Agent & Deliver $\rho_i$ and finalise success \\
\makecell[l]{\textsc{Report}\\\textsc{-Failed}} & Agent & Deliver typed failure and finalise error \\
\textsc{Heartbeat} & Agent & Refresh liveness for owned tasks \\
\midrule
\textsc{Requeue} & Admin & Return stale running task to \Queued \\
\textsc{Force-Fail} & Admin & Move non-terminal task to \Failed \\
\bottomrule
\end{tabular}
\end{table}

A key separation is between \textsc{Claim} and \textsc{Report-Running}. \textsc{Claim} is the control-plane commitment step: it establishes $\omega(\tau_i)$ and transfers the execution contract $(\tau_i, \delta_i)$, but does not yet prove that the scheduler has accepted the job. \textsc{Report-Running} marks the realisation step: the task has crossed into scheduler-backed execution and now carries concrete batch metadata. This separation localises failures precisely: claim failure, submission failure, runtime failure, and result-finalisation failure occur at distinct protocol points and can be diagnosed independently. The heartbeat is intentionally weak: it does not prove fine-grained forward progress. Its role is to establish that an owner is still present and aware of its ongoing tasks, which is sufficient for bounded recovery. When ownership remains unreaffirmed beyond the configured liveness window, operators can re-queue the task or explicitly force-fail it.

\textbf{Invariant D4.} Every task in \Running either receives reaffirming report or heartbeat traffic from its recorded owner within the configured liveness window, or becomes eligible for explicit recovery. \hfill\emph{(discharges C6)}

\subsection{D5: Device Snapshot as Immutable Execution Contract}
\label{sec:design-snapshot}

Let $\tau_i=(c_i,\ell_i,s_i,d_i)$. C3 requires that the result be determined by the bound execution contract, which we write as
\[
\rho_i = \mathcal{E}(\tau_i, \delta_i),
\qquad
\delta_i = \Delta(\mathcal{D}_{d_i}, t_{\mathrm{claim}}(\tau_i)).
\]
That result must remain invariant with respect to any subsequent evolution of $\Delta(\mathcal{D}_{d_i}, t)$ for $t > t_{\mathrm{claim}}(\tau_i)$. The design tension is when to bind $\delta_i$. Binding at submission time is too early: the interval between submission and actual execution may span minutes to hours of scheduler queueing, during which the device state $\Delta(\mathcal{D}_{d_i}, t)$ may evolve through calibration refreshes or availability changes. Binding at compute-node runtime is too late: it creates a live network dependency from the scheduled job back to the control plane, violating execution hermeticity and introducing failure modes inside the most constrained phase of the system. \sys resolves this by binding $\delta_i$ at \emph{claim time}. The \textsc{Claim} response carries the full snapshot, which the agent materialises into the local task payload. Claim thereby transforms $\tau_i$ from a logically stateless API object into a time-indexed, immutable execution contract: once $\omega(\tau_i)$ is established, the owning agent also possesses the exact topology, calibration, and noise parameters that will govern execution. Scheduler latency may postpone execution, but cannot silently rewrite the contract under which the task runs.

\textbf{Invariant D5.} A scheduled job can determine its full execution behaviour from its local payload alone: $\rho_i = \mathcal{E}(\tau_i, \delta_i)$ requires no live control-plane access. \hfill\emph{(discharges C3)}

\subsection{D6: Topology-Parametric Validation and Execution}
\label{sec:design-topology}
Admissibility checking $c_i \vdash \Delta(\mathcal{D}_{d_i}, t)$, defined in Section~\ref{sec:background}, requires verifying that every qubit referenced in $c_i$ belongs to $Q$, every gate belongs to $\mathcal{G}_{\mathrm{native}}$, and every multi-qubit operation acts on a coupling in $E$. A naive implementation might branch on vendor or topology family, but this would be brittle to extend, awkward for hypothetical architectures, and contrary to the graph-structured generality of $\Delta$. \sys instead treats topology and calibration uniformly as data. Both admission-time validation and execution-time interpretation consult the same snapshot-defined graph, gate set, and calibration fields; no topology family receives bespoke control logic. A hypothetical processor is therefore represented in the same way as a measured one: as a snapshot whose structure and parameters have been constructed rather than imported.

\textbf{Invariant D6.} Validation and execution are parameterised by snapshot data, not by topology-specific code branches. \hfill\emph{(supports C3)}

\subsection{D7: TTL-Governed Snapshot Caching}
\label{sec:design-cache}

The device snapshot $\Delta(\mathcal{D}_d, t)$ lies on three performance-critical paths: admissibility checking at submission, snapshot binding at claim, and device-listing queries. The tension is therefore between read latency and mutation visibility. Unbounded caching would accelerate reads but risk admissibility checks passing against a stale snapshot or claims binding obsolete calibration into execution. \sys resolves this with a TTL-governed cache for read-heavy access paths together with immediate invalidation on administrative mutation. The detailed cache policy is deferred to Section~\ref{sec:impl-layers} and Appendix~\ref{supp-app:design-cache}.

\textbf{Invariant D7.} Let $\mu_d$ denote an administrative mutation to device $\mathcal{D}_d$ committed at time $t_\mu$. Any subsequent read of $\mathcal{D}_d$ that is ordered after $t_\mu$ returns a snapshot consistent with $\mu_d$. Reads not ordered after an administrative mutation may be served from cache, but the returned snapshot may lag the authoritative device state by at most the configured TTL. This bounded staleness does not affect execution-contract correctness, because execution snapshots are bound explicitly at claim time. \hfill\emph{(supports C3)}

\subsection{D8: Hermetic Compute-Node Execution}
\label{sec:design-runner}
C2 requires that every $\tau_i$ execute under a scheduler-allocated job $j_i \in \mathcal{J}$; C3 together with D5 require that execution depend only on the local payload $(\tau_i, \delta_i)$. These constraints jointly demand that the compute-node runner be hermetic: it must neither reach back to the control plane nor import the service runtime. If the runner were allowed to make live API calls or use mutable shared state, the compute phase would become a distributed service with a larger blast radius and worse reproducibility. \sys therefore enforces a strict local boundary: the runner consumes the bound task payload, performs simulation via $\mathcal{E}$, and emits a local result artifact $\rho_i$. The login-node agent is the sole bridge between local filesystem state and remote service state; the concrete runner pipeline is deferred to Section~\ref{sec:impl-runner}.

\textbf{Invariant D8.} The compute-node execution path is self-contained: given its run directory, $\rho_i = \mathcal{E}(\tau_i, \delta_i)$ is replayable without live control-plane or device-registry dependencies. \hfill\emph{(discharges C2, reinforces C3)}

\subsection{D9: Service-Side Observability}
\label{sec:design-sse}

C6 requires observable task progress and bounded recovery; D1 prohibits new inbound paths into the HPC environment. Together, they create a tension specific to the two-plane architecture: task execution unfolds asynchronously within a domain the client cannot reach, yet the client must track lifecycle state with sufficient timeliness to act on failures, coordinate dependent workflows, or present meaningful progress.

The design principle is that observability must be derived entirely from service-side state, not from execution-side instrumentation. The control plane already records every authoritative state transition: submission, claim, completion, failure, cancellation, and device mutation. Observability is therefore a projection of these committed events to external consumers, not a new sensing path into the cluster. \sys realises this principle through a server-initiated event stream from the control plane; the concrete bounded-memory fan-out mechanism is deferred to Section~\ref{sec:impl-sse} and Appendix~\ref{supp-app:design-sse}.

\textbf{Invariant D9.} Clients can observe task and device state changes in near real-time; the observation path is derived entirely from control-plane state and requires no communication into the HPC environment. \hfill\emph{(discharges C6, compatible with D1)}

\subsection*{Design Summary}

The nine invariants compose into a single architectural result. D1 restricts cross-boundary communication to outbound initiation; D3 and D4 then maintain exclusive ownership and bounded liveness through an atomic claim and a narrow asymmetric protocol. D5 converts the owned task into an immutable execution contract by binding $\delta_i$ at claim time; D6 and D7 keep that snapshot topology-parametric and freshness-bounded; D8 guarantees that the compute-node runner executes hermetically from this contract alone. D2 and D9 project the resulting asynchronous execution back into a stable, observable service abstraction. The net effect is a clean separation between service authority and scheduler-governed realisation, with no shared process state, storage, or inbound connectivity.

Table~\ref{tab:invariant-map} maps each invariant to the implementation mechanism that enforces it and the experiment that exercises it, so that each later claim can be traced back to its design origin.

\begin{table}[!htp]
\centering
\caption{Design invariants, the implementation mechanism that enforces each, and the experiment that exercises it. Section~\ref{sec:impl} develops the mechanisms; Section~\ref{sec:eval} reports the evidence.}
\label{tab:invariant-map}
\footnotesize
\setlength{\tabcolsep}{3.5pt}
\renewcommand{\arraystretch}{1.05}
\begin{tabularx}{\columnwidth}{@{}l Y Y@{}}
\toprule
Inv. & Mechanism (\S\ref{sec:impl}) & Evidence (\S\ref{sec:eval}) \\
\midrule
D1 & Capability-partitioned two-artifact deployment; outbound-only protocol & Bounded service overhead; outbound-only production run \\
D2 & Authoritative task store with static transition relation & Lifecycle preserved through crash and recovery \\
D3 & Atomic \textsc{Claim} as ownership linearisation point & 50/50 tasks claimed exactly once under two agents \\
D4 & Heartbeat expiry plus reconciliation-driven recovery & 3/3 \Running\ tasks recovered after agent crash \\
D5 & Claim-time snapshot binding via \texttt{SimulatorPort} & 8/8 tasks reflect post-claim mutation, not stale state \\
D6 & Graph-parametric admissibility over snapshot topology & Calibration-aware run on 156-qubit heavy-hex device \\
D7 & TTL cache with invalidate-on-mutate for device reads & Post-mutation reads visible immediately at claim \\
D8 & Artifact-local execution sealed to bound contract & Runner reproduces $\rho_i=\mathcal{E}(\tau_i,\delta_i)$ from local payload \\
D9 & Authoritative event projection with bounded-memory SSE & Latency reconstructed from authoritative event stream \\
\bottomrule
\end{tabularx}
\end{table}

\section{Implementation}
\label{sec:impl}

This section describes how \sys realises the invariants from Section~\ref{sec:design}. The implementation spans three deployment domains with different failure characteristics: a cloud-hosted control plane subject to concurrent client access, a login-node agent operating under scheduler latency and transient connectivity, and ephemeral compute nodes that are network-isolated from the service. The body focuses on the implementation boundaries that are load-bearing for the architecture; detailed admissibility logic, noise-model construction, fault provenance, and the invariant-to-implementation synthesis are deferred to Appendix~\ref{supp-app:impl-details}--Appendix~\ref{supp-app:impl-map}.

\subsection{Two-Artifact Deployment}
\label{sec:impl-isolation}

\sys is implemented as two independently deployable artifacts: \texttt{vqpu\_server} and \texttt{vqpu\_agent}. The split is a capability boundary. The server artifact contains task admission, lifecycle management, device-state management, snapshot binding, authentication, event projection, and result ingestion. It contains no scheduler-facing functionality, SSH dependency, cluster-addressing module, or compute-side launcher. Conversely, the agent artifact contains the site-side launch path and outbound protocol client, but no public listener and no device-management API.

This partition makes D1 a property of construction rather than operator discipline. The control plane cannot reach into the cluster because it lacks the capability, while the agent needs only the privileges of an ordinary user process on a login node. Deployment therefore preserves the site's existing security posture: no firewall exception, privileged daemon, or administrator-managed inbound service is required. The split also supports operational independence. The public API can evolve without editing Slurm templates, while site-side launch configuration can change without altering the client-visible service contract.

\subsection{Service-Layer Contract Boundaries}
\label{sec:impl-layers}

The control plane uses a ports-and-adapters architecture~\cite{cockburn2005hexagonal}. Domain services manipulate tasks, device snapshots, domain events, structured results, and typed failures through abstract ports. Concrete adapters, including SQLite stores, an in-memory cache, JSONL event storage, authentication, and simulator backends, are attached at the composition root. The purpose is not only modularity: it prevents scheduler state, storage-specific behaviour, or backend-local runtime state from entering the service semantics accidentally.

The port layer is organised around three contract families. The first governs lifecycle and identity: task storage, event persistence, authentication, and principal information. The second governs device state: authoritative device access, cached snapshot views, and mutation-triggered invalidation. The third governs execution: a simulator interface that consumes the bound task contract. This organisation makes the dependency structure mirror the paper's architecture. Control-plane logic cannot accidentally depend on Slurm state because no such dependency is present in the domain core.

\texttt{SimulatorPort} is the critical execution boundary. Its signature accepts only the execution-relevant components of a task: circuit $c_i$, dialect $\ell_i$, shot count $s_i$, and bound snapshot $\delta_i$, and returns a structured result $\rho_i$. There is no interface by which the service can consult mutable backend-local device state or scheduler state. This is what makes D5 concrete in code: from the control plane's perspective, execution is parameterised by the bound contract. It also supports D6, because the same interface handles a 20-qubit test graph, a 156-qubit heavy-hex snapshot, or a synthetic future topology without changing control-plane logic.

\texttt{DeviceStorePort} implements the freshness boundary. Administrative mutations operate on authoritative device state and immediately invalidate cached views. User-facing listing, viability checking, and other read-heavy paths may use a TTL-governed snapshot view. This gives low-latency interaction without weakening execution correctness: the admissibility path may use a bounded-staleness view, but the execution contract is bound from authoritative state at claim time. Appendix~\ref{supp-app:impl-viability} expands the two-stage admissibility path, and Appendix~\ref{supp-app:impl-noise} details the snapshot-derived noise model used by both service-side and compute-side execution paths.

Submission-time admissibility is deliberately lightweight. The service performs a syntactic, graph-parametric check over the submitted circuit to reject requests that reference unavailable qubits, unsupported native operations, or illegal couplings. This early check avoids wasting scheduler resources and gives immediate diagnostics to the caller, but it is not the final execution contract. The final contract is the claim-time snapshot. This two-stage structure keeps the API interactive while preserving the stronger semantic guarantee that execution is governed by the state current at ownership transfer.

The same boundary is used for noisy execution. The snapshot is not only a topology descriptor; it is the complete simulation contract for the virtual QPU at a point in time. The simulator derives noise channels directly from the calibration fields carried by $\delta_i$, so two tasks with identical circuits but different snapshots may legitimately produce different output distributions. Because the control-plane simulator and compute-node runner use the same derivation logic, the meaning of the bound contract is independent of where the simulation is realised. This prevents a subtle but important failure mode: a task should not change semantics merely because it crossed from a service-side validation path to an HPC-side execution path.

\subsection{Atomic Claim and Authoritative Task Store}
\label{sec:impl-claim}

The control plane maintains, for each task $\tau_i$, an authoritative lifecycle state $\sigma(\tau_i)$ and owner identity $\omega(\tau_i)$. D2 and D3 are enforced by the same mechanism: a serialised task store whose linearisation point is the atomic \textsc{Claim} operation. A claim transaction selects an eligible queued task, rechecks its state within the same serialisation boundary, commits $\Queued\rightarrow\Running$, and records the claimant as owner. Competing agents cannot both acquire the same task, and expensive scheduler work begins only after ownership is uniquely established.

All later lifecycle evolution is constrained by the transition relation over the five-state alphabet in Eq.~\ref{eqn:task-lifecycle}. The store does not expose arbitrary mutation of task state; it admits only legal transitions and rejects invalid ones with typed errors. Terminal states are absorbing, including for administrative operations. Scheduler or runner observations may provide evidence, but only authenticated protocol actions through the task-store boundary change service truth. Each committed transition is also appended to the event log, giving the system both an authoritative present state and an auditable history.

This structure cleanly separates authority from evidence. The scheduler may indicate that a job started, exited, or failed; the runner may produce a result artifact; the agent may observe stale local directories after restart. None of these observations directly defines $\sigma(\tau_i)$. They become service-visible only by passing through a legal transition. This separation is what lets \sys preserve a compact client-facing lifecycle while execution traverses a richer and site-specific scheduler state machine underneath.

The same boundary handles administrative intervention. Operations such as requeue and force-fail are represented as explicit lifecycle transitions rather than privileged writes around the state machine. This is important for auditability: recovery does not bypass the service contract, it re-enters it through a typed transition whose preconditions can be checked and whose effect can be recorded as an event.

\subsection{Agent Reconciliation and Finalisation}
\label{sec:impl-agent}

The agent is a reconciliation controller rather than an authoritative execution database. It observes three partially overlapping sources: service-side task state, scheduler or adapter evidence, and local run-directory artifacts. Its filesystem records materialised work, bound snapshots, launch metadata, and result artifacts, but not task truth. After restart, recovery is therefore observational: the agent re-reads the control plane, scheduler evidence, and local directories, then computes the minimal action needed to restore consistency.

This design separates execution from publication. A scheduler job may complete, fail, be rediscovered, or be observed repeatedly, but publication of a terminal outcome is guarded by the control-plane transition boundary. Once a task has reached \Completed, \Failed, or \Cancelled, no later observation can induce a second terminal transition. In operation, the agent multiplexes three bounded pipelines: work acquisition, heartbeat maintenance, and artifact finalisation. Heartbeats assert owner presence and awareness of owned work; they do not claim fine-grained forward progress. Finalisation inspects the result artifact, normalises success or failure, and attempts exactly one legal terminal publication.

The resulting failure behaviour is conservative by design. If an agent crashes after claiming a task, the control plane still records the last authoritative fact: the task is \Running\ and owned by that agent. A later heartbeat expiry makes the task eligible for explicit recovery, but it does not itself prove that the scheduler job failed. This is the implementation counterpart of D4: liveness absence is treated as recoverable evidence, not as a hidden automatic state transition.

\subsection{Scheduler Integration Boundary}
\label{sec:impl-slurm}

The scheduler boundary is deliberately narrow:
\[
\mathcal{B}_{\mathrm{sched}}=\{\textsc{Submit},\textsc{ObserveActive},\textsc{QueryTerminal}\}.
\]
The agent needs only to submit a rendered job, determine whether owned work is still active, and recover terminal accounting evidence once a job leaves the active set. It does not require full scheduler programmability, direct mutation of scheduler objects, or global queue introspection.

The current Pawsey deployment realises this boundary through a Dask-backed Slurm adapter. Dask is used as an execution adapter, not a replacement for the site scheduler: resource allocation remains scheduler-mediated. Partition names, account strings, walltime, modules, environment activation, container wrappers, filesystem conventions, and Dask cluster parameters all live below this boundary in site-local configuration. Above it, task claiming, snapshot binding, runner semantics, and lifecycle finalisation remain scheduler-adapter agnostic.

This is the narrow waist that makes \sys portable in the architectural sense. Replacing Slurm with PBS Pro, LSF, Flux, or a Kubernetes-style batch backend would require a different realisation of $\mathcal{B}_{\mathrm{sched}}$, but not a different task lifecycle, device model, snapshot contract, or external API. Conversely, changing the virtual-device model or REST API does not require rewriting the site-specific launch wrapper unless the execution payload itself changes.

\subsection{Hermetic Compute-Node Runner}
\label{sec:impl-runner}

The compute-node runner realises the local transformation
\[
(\tau_i,\delta_i)\longrightarrow\rho_i
\]
from a run directory containing the bound task payload and device snapshot. It opens no network connection, consults no live device registry, imports no control-plane module, and writes only a local result artifact. It parses the submitted circuit, constructs the simulator context and noise model from $\delta_i$, transpiles against the snapshot-defined coupling map and native gates, runs Qiskit-Aer/cuQuantum, and writes structured output.

This boundary is load-bearing for both reproducibility and fault containment. Re-running the preserved directory reconstructs the same virtual-device semantics even if live device state has changed. Conversely, a backend exception, node crash, or malformed result cannot directly corrupt service state because the runner has no write path into the control plane. The agent must translate local evidence into one legal lifecycle action.

Hermeticity also makes claim-time binding operationally meaningful. If the runner were allowed to consult mutable device state during execution, then the claim-time snapshot would be advisory rather than authoritative. By closing the compute path over the local payload, \sys ensures that later device mutations, cache refreshes, or server outages cannot retroactively change the semantics of an already-claimed task.

\subsection{Bounded-Memory Event Projection}
\label{sec:impl-sse}

The SSE stream implements D9 as a projection of committed control-plane state. Events are emitted from authoritative actions such as submission, claim, completion, failure, cancellation, and device mutation. They are not a second source of truth and not a new observation path into the cluster.

The broker uses bounded per-subscriber queues and a replay window. Slow consumers are disconnected rather than allowed to block publication; reconnecting clients can recover missed history. Publications from synchronous callbacks are routed onto the broker's event loop, preserving serialisation at the projection boundary without introducing coarse-grained locking.

This event design keeps observability from becoming an execution dependency. A dashboard may disconnect, a browser may stall, or a subscriber may fall behind, but task publication and runner reporting continue. Conversely, all events remain derived from committed control-plane state, so clients do not need to combine scheduler logs, agent logs, and REST polling to reconstruct task meaning.

\subsection*{Summary}
\label{sec:impl-summary}

The implementation realises \sys through explicit boundaries: capability partitioning enforces outbound-only connectivity; ports-and-adapters preserve snapshot-parameterised semantics; the task store and agent provide authoritative start/finish control; the scheduler adapter localises site variability; the runner seals execution to the bound contract; and SSE projects service truth without coupling observation load back into execution.


\section{Evaluation}
\label{sec:eval}

We evaluate whether \sys preserves the semantic and operational properties required to export an HPC-resident simulator as an interactive, device-aware service. The evaluation targets five claims: (i) service-export overhead is bounded and does not scale with circuit size; (ii) device snapshots carry execution-relevant calibration semantics; (iii) claim-time binding prevents stale-snapshot execution when device state changes before claim; (iv) atomic claim preserves task integrity under concurrent agents; and (v) authoritative lifecycle state supports explicit recovery after agent failure.

The evidence has two layers. A deterministic regression suite exercises lifecycle legality, duplicate completion rejection, snapshot-cache behaviour, SSE ordering and replay, topology-parametric validation, and related invariants. End-to-end experiments then exercise the same properties on the Pawsey deployment under real scheduler delay, GPU execution, and device snapshots. Unless otherwise stated, execution uses \texttt{qiskit-ella}, which invokes Qiskit-Aer with cuQuantum on Setonix GPUs. The primary target is \texttt{ibm-fez-0309}, a 156-qubit virtual device instantiated from IBM Fez calibration data captured on 2026-03-09; \texttt{ibm-fez-ideal} preserves the same topology and native gate set with null error fields. Full setup and supplementary controls are in Appendix~\ref{supp-app:eval-method}--\ref{supp-app:eval-summary}.

The experiments are intentionally property-driven rather than benchmark-only. The goal is not to claim that \sys makes simulation faster than direct scheduler use; the simulator and scheduler remain the performance substrate. The goal is to test whether the service layer preserves device semantics and lifecycle authority while adding only bounded orchestration cost.

\subsection{Bounded Service-Path Overhead (D1, D4, D9)}
\label{sec:latency}

This experiment asks whether \sys introduces a workload-dependent scaling penalty or only a bounded service-export cost. We submit random native-gate circuits with $n\in\{28,29,30,31,32\}$ qubits, 1024 shots, and five repetitions per size. Each circuit contains 10 layers of random single-qubit gates followed by a random maximal matching of CZ gates over the device graph. End-to-end latency is decomposed as
\[
T_\mathrm{e2e}=T_\mathrm{admit}+T_\mathrm{queue\mbox{-}claim}+T_\mathrm{exec}+T_\mathrm{poll},
\]
where admission is measured at the client and the remaining terms are reconstructed from authoritative lifecycle timestamps. Runner instrumentation further decomposes $T_\mathrm{exec}$ into parsing, noise construction, transpilation, and simulation.

The circuit family is intentionally structure-free. Its purpose is not to benchmark a particular algorithm, but to make the workload-dependent component primarily state-vector simulation. If the service layer introduced a second scaling law, it would appear as increasing admission, claim, snapshot, or event cost as $n$ grows. If \sys behaves as intended, those costs should remain bounded while the simulator term grows.

\begin{figure*}[!htp]
  \centering
  \subfloat[End-to-end latency decomposition.\label{fig:latency-e2e}]{%
    \includegraphics[width=0.48\textwidth]{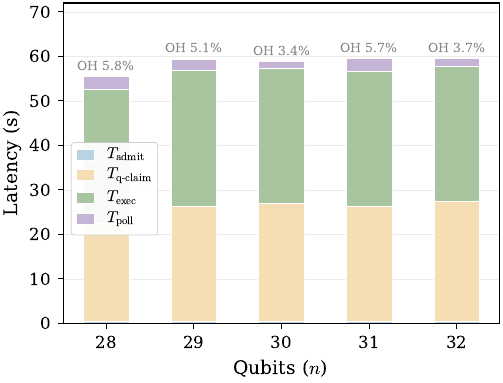}}
  \hfill
  \subfloat[Execution-phase breakdown.\label{fig:latency-exec}]{%
    \includegraphics[width=0.48\textwidth]{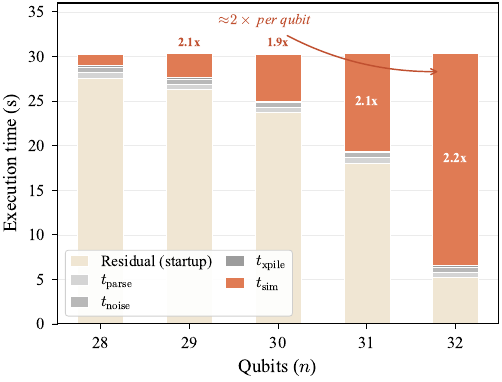}}
\caption{Latency decomposition across 28--32 qubits on \texttt{ibm-fez-0309} (1024 shots, mean of 5 trials). Interface-visible overhead remains 3.4--5.9\% of end-to-end latency; runner-side simulation grows at approximately $2\times$ per added qubit while parse, noise construction, and transpilation remain nearly constant. Exact values are in Tables~\ref{supp-tab:latency} and~\ref{supp-tab:sim-breakdown}.}
  \label{fig:latency}
\end{figure*}

Figure~\ref{fig:latency-e2e} shows that $T_\mathrm{admit}$ and $T_\mathrm{poll}$ remain effectively constant across all qubit counts. The dominant non-compute term is $T_\mathrm{queue\mbox{-}claim}$, which lies between 22\,s and 27\,s and reflects agent polling and scheduler interaction rather than circuit size. Figure~\ref{fig:latency-exec} shows the expected compute-side trend: simulation rises from 1.32\,s at 28 qubits to 23.77\,s at 32 qubits, with per-qubit ratios close to $2\times$, while parsing, noise construction, and transpilation remain essentially constant.

The important result is separation of scaling behaviour. The fixed service and startup envelope is additive: at 28 qubits it dominates the short simulation, while by 32 qubits the GPU simulation term accounts for 78.4\% of runner time. \sys therefore does not hide the exponential cost of state-vector simulation; it prevents that cost from leaking into admission, ownership, snapshot handling, or event projection. Appendix~\ref{supp-app:eval-latency} gives the full numerical tables and scaling-ratio analysis.

This distinction is important for interpreting the nearly flat end-to-end latency across 28--32 qubits. The flatness is not evidence that the underlying computation is constant. It shows that the experiment operates near the crossover between a fixed orchestration floor and the exponential simulator term. As circuit size grows, end-to-end latency must eventually track the simulator. The architectural claim is therefore narrower and stronger: \sys preserves the scaling behaviour of the underlying HPC simulator rather than introducing an additional workload-dependent service term.

The same decomposition also explains why the overhead is acceptable for the target use case. Interactive virtual-device access does not require that every short circuit run with zero service cost; it requires that the service path be predictable, bounded, and independent of the scientific workload once a task has been admitted. The measured interface-visible overhead is 3.4--5.9\% of end-to-end latency in this experiment, while the absolute API-facing components remain small relative to the scheduler-mediated path. For larger simulations, parameter sweeps, and calibration studies, the fixed service envelope is amortised by the compute workload. For very small circuits, the result makes the tradeoff explicit: \sys buys device semantics, lifecycle authority, and remote observability at the cost of a bounded orchestration floor.

\subsection{Device-Aware Simulation Fidelity (D5, D6)}
\label{sec:eval-noise}

This experiment checks whether the snapshot is executable semantics rather than passive metadata. We use an amplified two-qubit identity workload whose ideal output is deterministically $\ket{00}$. The two conditions share topology, native gate set, validation path, claim path, and GPU backend; only calibration content differs. \texttt{ibm-fez-ideal} carries null calibration, while \texttt{ibm-fez-0309} carries real Fez calibration. Each condition uses 8192 shots and five trials.

Operationally, the workload prepares a simple entangling pattern and repeatedly applies the calibrated two-qubit channel on edge $(0,1)$. This amplifies the effect of edge and readout calibration without changing topology or routing. The experiment therefore isolates the causal variable: if output distributions differ, they differ because the bound snapshot changes the noise model instantiated by the runner.

\begin{figure}[!htp]
  \centering
  \includegraphics[width=0.85\columnwidth]{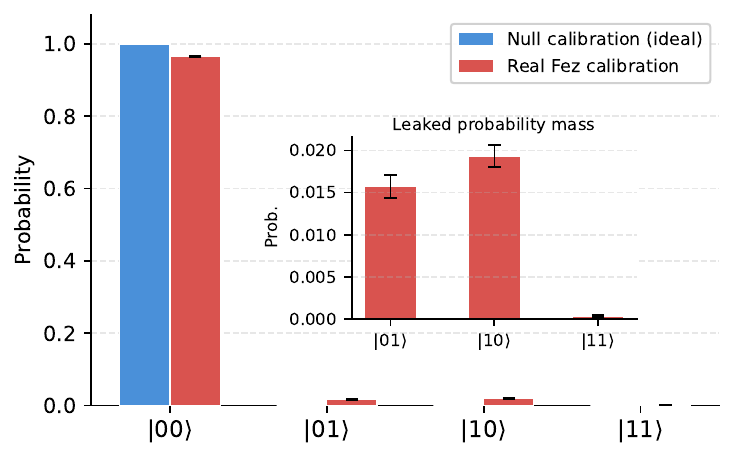}
  \caption{Output-state probability distribution for the amplified two-qubit identity circuit under null-calibration and real-calibration snapshots (8192 shots, mean of 5 trials). The inset magnifies non-$\ket{00}$ states; error bars denote one standard deviation.}
  \label{fig:noise-dist}
\end{figure}

Figure~\ref{fig:noise-dist} shows a clear calibration effect. Under the null snapshot, only $\ket{00}$ is observed. Under the real Fez snapshot, the mean total-variation distance from the ideal distribution is $TV=0.03537$ with standard deviation 0.00171. The leakage is structured: probability mass is concentrated primarily on $\ket{10}$, consistent with non-uniform readout error and the calibrated two-qubit channel on edge $(0,1)$. Because the circuit, topology, API path, claim path, and GPU backend are fixed, the output shift is attributable to calibration content in the bound snapshot. This also exercises graph-parametric validation and noise construction on a 156-qubit heavy-hex snapshot; Appendix~\ref{supp-app:eval-snapshot} gives additional interpretation.

The result rules out a weak interpretation of the snapshot as documentation attached to a task. The snapshot changes the simulator state that the runner constructs and therefore changes the measurement distribution, providing the experimental link between the design contract in D5 and the requirement that calibration data remain part of execution semantics. The two-qubit circuit makes the ideal reference unambiguous, while reusing the 156-qubit Fez snapshot exercises the same validation and reconstruction path used by larger tasks.

\subsection{Claim-Time Snapshot Binding (D5, D7)}
\label{sec:eval-binding}

The fidelity experiment establishes that calibration content affects execution. The binding experiment identifies when that content becomes fixed. We submit eight tasks to \texttt{ibm-fez-0309} while it carries real calibration. Approximately 1.2\,s later, before the agent claims the queue, we update the same device through the administrative API and set all noise parameters to zero. The claim occurs later, after the mutation has become current. Submit-time binding predicts the noisy fingerprint from Section~\ref{sec:eval-noise}; claim-time binding predicts exact output.

The timing creates a direct counterfactual for the same device identity. If tasks capture device state at submission, the later administrative mutation should not matter. If tasks bind at claim, the mutation should change the execution contract for the entire batch. If tasks perform a late live lookup from the compute node, the result would depend on runtime reachability and would violate hermetic execution. The expected \sys behaviour is the second case.

\begin{figure}[!htp]
  \centering
  \includegraphics[width=0.85\columnwidth]{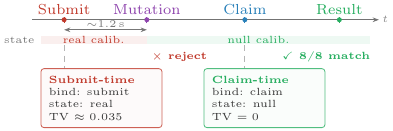}
  \caption{Same-device temporal separation test for snapshot binding. Eight tasks are submitted while \texttt{ibm-fez-0309} is noisy; before claim, the same device is administratively zeroed. The observed outcome matches claim-time binding: all eight tasks execute with the post-mutation ideal snapshot.}
  \label{fig:binding-scene-b}
\end{figure}

All eight tasks return $TV=0$ and $p(\ket{00})=1.0$, so none retains the noisy fingerprint expected under submit-time binding. Freshness is therefore resolved once, at ownership transfer, and the resulting snapshot is carried immutably into scheduler-backed execution. Appendix~\ref{supp-app:binding-scene-a} reports the cross-device interleaving control, which rules out misbinding between noisy and ideal device identities.

Because the same device identity appears on both sides of the mutation, the only difference between the two predictions is the lifecycle point at which the snapshot is bound: the observed all-ideal outcome therefore identifies that point as claim, not submission and not compute-node runtime. Operationally, this matches HPC queue reality: freshness is decided as late as possible before ownership begins, while execution itself does not depend on live control-plane reachability.

\subsection{Crash Recovery and Authority (D2, D3, D4)}
\label{sec:eval-recovery}

The preceding experiments test device semantics. This experiment tests authority when the execution side becomes unreliable. We submit three tasks, wait until the agent claims them and transitions them to \Running, and then terminate the agent process on the Pawsey login node. The control plane preserves all three tasks in \Running; it does not auto-fail, auto-timeout, or auto-requeue them based solely on loss of observation. We then issue one administrative requeue per task, restart the agent, and observe all three tasks complete.

This behaviour is intentionally conservative. In an HPC setting, stale liveness is ambiguous: it may reflect agent failure, scheduler delay, node disruption, or partial loss of observability. \sys therefore records what has been authoritatively committed and leaves ambiguity explicit until an authorised recovery action resolves it. The recovery record, 3/3 completed after one requeue per task, shows that recovery is bounded, auditable, and re-enters the normal claim path without forking execution history. Appendix~\ref{supp-app:eval-ops} expands this operational interpretation.

The key evidence is not merely that the tasks eventually finish. It is that the control plane does not silently reinterpret missing observations as failure or success. A more aggressive system could auto-requeue on heartbeat expiry, but that would risk competing execution lineages if the original scheduler job later produced a result. \sys instead keeps ambiguity visible and requires explicit recovery, which is the safer semantic choice across an HPC boundary.

\subsection{Multi-Agent Concurrency (D3)}
\label{sec:eval-concurrency}

This experiment tests whether atomic claim remains correct when execution authority is distributed. We submit 50 random 5-qubit circuits in a burst while two independent agents run concurrently: \texttt{ella-runner-1} and \texttt{setonix-runner-1}, each configured with two scheduler-backed execution slots.

The experiment separates three possible outcomes. The first is semantic failure: duplicate claims, ambiguous ownership, or duplicate completion. The second is correctness without useful concurrency: all tasks complete once, but coordination collapses throughput well below the four available slots. The desired outcome is both semantic integrity and near-capacity utilisation without a central dispatcher.

\begin{table}[h]
  \centering
  \caption{Multi-agent concurrency under two independent runners, each exposing two concurrent execution slots.}
  \label{tab:concurrency}
  \small
  \begin{tabular}{lrrr}
    \toprule
    Runner & Tasks & Share & Mean $T_\text{e2e}$ (s) \\
    \midrule
    \texttt{ella-runner-1}    & 24 & 48.0\% & 201.5 \\
    \texttt{setonix-runner-1} & 26 & 52.0\% & 211.3 \\
    \midrule
    Total                     & 50 & 100\%  & --- \\
    \bottomrule
  \end{tabular}
\end{table}

All 50 tasks reach \Completed with no failures, timeouts, duplicate completions, or ambiguous ownership. The split is near even at 24:26. The batch finishes in 402.9\,s, only about 3.3\% above a simple four-slot lower bound of 13 execution waves at roughly 30\,s per wave. The result is not merely a throughput observation: it shows that the queue plus atomic claim acts as a decentralised arbitration surface. Agents coordinate at the smallest semantically necessary boundary, while execution remains local, asynchronous, and scheduler-mediated.

This result closes the loop on D3 under realistic contention. In a single-agent deployment, atomic claim is primarily a correctness mechanism. With two agents, it also becomes the system's arbitration primitive. The near-even split and near-slot-limited completion time show that exclusivity is not purchased by serialising all execution through a central dispatcher.

\subsection{Summary}
\label{sec:eval-summary}

The evaluation validates the central service-export claim. Interface overhead is bounded and additive; calibration-bearing snapshots materially affect GPU output; claim-time binding chooses the effective device state after queue delay; concurrent agents complete 50/50 tasks exactly once; and explicit recovery restores 3/3 stale \Running tasks without silent lifecycle reinterpretation. Appendix~\ref{supp-app:eval-summary} links each result back to the design invariants.

\section{Related Work}
\label{sec:related}

The work most relevant to \sys spans several communities: high-performance quantum simulation, quantum cloud platforms, quantum--HPC middleware, HPC job and workflow services, pull-based worker systems, and quantum resource virtualisation. These areas provide important ingredients for performance, APIs, orchestration, and runtime management. None directly addresses the system boundary studied here: exporting device-aware quantum simulation as an interactive service from an inward-closed, batch-scheduled HPC environment.

\subsection{High-Performance Quantum Simulation}

Qiskit-Aer, cuQuantum, SV-Sim, and related systems address simulation performance, scalability, and backend portability across heterogeneous CPU/GPU resources~\cite{JavadiAbhari2024,bayraktar2023cuquantum,Li2021}. SV-Sim is particularly relevant because it identifies irregular remote communication in distributed state-vector simulation as a precise HPC bottleneck~\cite{Li2021}. \sys is complementary. It does not propose a new kernel or communication substrate; it asks how an existing HPC-hosted simulator can be exported as a device-aware service despite scheduler mediation and restricted reachability.

\subsection{Quantum Cloud Platforms}

IBM Quantum, Amazon Braket, Google Quantum AI, and related services establish the device-oriented abstraction that motivates \sys: named backends, structured circuit submission, managed results, and programmatic APIs~\cite{IBMQuantum2026,AWS2025,cirq_google}. In those platforms, however, the operator typically controls both service boundary and execution substrate. \sys targets the case in which execution occurs inside an independently administered HPC site with its own scheduler, security policy, and software environment. The contribution is therefore not an API alone, but cloud-like device semantics over an execution substrate the service cannot directly control.

\subsection{Quantum--HPC Middleware and Workflow Systems}

Quantum--HPC middleware work identifies integration patterns and builds modular frameworks for hybrid execution across classical and quantum resources~\cite{Saurabh2023,Shehata2024,Shehata2026,Mantha2025}. HPC portals and workflow systems such as Globus, Pegasus, Parsl, Dask, and Galaxy provide remote submission, orchestration, data movement, and workflow abstraction~\cite{pawsey_vqpu_hybrid_workflow,foster2006globus,deelman2015pegasus,babuji2019parsl,rocklin2015dask,galaxy2022galaxy}. These systems are relevant but usually centre the job, workflow, dataset, or resource reservation. \sys centres a graph-structured, calibration-aware virtual device. Once the service object is a device rather than a job, the core problem becomes freshness of calibration state, claim-time binding, hermetic execution, and lifecycle authority across scheduler delay.

This device-centred view changes the correctness question. For a workflow manager, a useful abstraction may be that a task graph eventually runs and produces files. For \sys, the user-facing result must remain interpretable as execution under a particular virtual-device contract. That contract includes admission against topology and native gates, selection of a calibration-bearing snapshot at a defensible time, and prevention of duplicate terminal outcomes under concurrent agents. Existing middleware can be valuable beneath or beside this architecture, but it does not by itself define these quantum-specific service semantics.

\subsection{Pull-Based Workers and Quantum Virtualisation}

Task queues such as Celery and RQ show how submission and execution can be decoupled through pull-based workers~\cite{celery_docs,rq_python}. \sys uses this intuition, but the similarity is limited: conventional queues assume a reachable broker, stable worker runtime, and in-process task execution, whereas \sys coordinates scheduler-mediated jobs on ephemeral compute nodes and preserves a time-indexed device contract. Quantum resource-virtualisation systems such as HyperQ, DynQ, and user-centric HPC-QC runtimes focus on resource sharing, qubit-level partitioning, runtime scheduling, or hybrid policy~\cite{tao2025quantum,liu2026dynq}. These directions are complementary: a virtual device produced by a future sharing-oriented runtime could be exported through \sys as another snapshot-defined target.

The relationship to worker systems also explains why exactly-once terminology must be used carefully. \sys does not rely on a broker delivering a message exactly once to a permanently reachable worker. It defines an authoritative lifecycle in which claim, ownership, and terminal publication are linearised at the control plane, while execution evidence is submitted from the HPC side. This is weaker than eliminating every possible physical duplicate execution under arbitrary infrastructure failure, but stronger and more relevant for the service boundary: clients observe one authoritative task history and one terminal result.

\subsection{Positioning}

Prior work provides fast simulators, device APIs, middleware patterns, workflow engines, worker coordination, and virtualisation concepts. What remains missing is a system design for the specific boundary condition addressed in this paper: interactive export of device-aware simulation from a secure, scheduler-mediated, inward-closed HPC environment. \sys occupies that position through a control/execution-plane split, outbound agent coordination, claim-time snapshot binding, hermetic runner execution, and scheduler-aware lifecycle semantics.


\section{Discussion and Conclusion}
\label{sec:discussion}

\subsection{Discussion}

The main contribution of \sys is not simply placing an API in front of a simulator. It identifies the semantic boundary required to export device-aware quantum simulation from secure HPC infrastructure. Quantum software expects a device-oriented abstraction: named backends, topology, calibration state, structured submission, and typed results. Production HPC exposes a scheduler-oriented abstraction governed by queueing, site policy, transient allocation, and inward-closed networking. The systems problem is to preserve the first abstraction while executing through the second.

\sys resolves this mismatch by separating service authority from scheduler-governed realisation. The control plane is authoritative for device identity, task lifecycle, snapshot semantics, and event projection. The execution plane realises that contract through claim, scheduler submission, hermetic compute-node execution, and artifact return. This separation is what allows the service to make device-aware promises before any HPC resource exists and still honour them after arbitrary scheduler delay.

The mechanisms developed in the paper are consequences of this boundary rather than isolated implementation choices. Outbound-only coordination makes service export compatible with HPC security. Atomic claim establishes a unique execution lineage when multiple agents race at the queue boundary. Claim-time snapshot binding reconciles calibration freshness with hermetic execution. Control-plane lifecycle authority prevents scheduler uncertainty or agent disappearance from silently rewriting task meaning. The evaluation shows these mechanisms operating together: overhead remains bounded and additive, calibration-bearing snapshots change output distributions, claim-time binding resolves queue-delay freshness, concurrent agents preserve ownership, and explicit recovery restores progress after failure.

The quantum-specific contribution is the exported contract. Because topology, native operations, and calibration state are first-class service inputs, \sys supports more than convenient remote execution. It lets simulation be consumed as a virtual-device service for compiler evaluation, routing analysis, topology exploration, calibration sensitivity, and hardware--software co-design. The same interface can expose current-device emulation and hypothetical future devices by changing snapshot structure and parameters.

This perspective also explains why the design invariants are not implementation embellishments. If outbound-only coordination is removed, the architecture no longer matches the security posture of many production HPC sites. If atomic claim is weakened, concurrent agents can create ambiguous ownership. If claim-time snapshot binding is replaced by live runtime lookup, execution semantics depend on compute-node reachability and mutable state. If hermetic runners are replaced by server-dependent jobs, scheduler delay and isolation become part of the scientific meaning of a result. The invariants therefore define the smallest service boundary that keeps the result interpretable as execution on a selected virtual QPU while still using ordinary HPC mechanisms.

Mechanisms such as outbound polling, explicit recovery, and hermetic contract binding are sometimes read as concessions to HPC, as if they approximated an ideal push-based cloud substrate. They are not. They are designed responses to a setting in which connectivity is asymmetric, execution is partially observable, and administrative authority is distributed across the site and the service. Under those constraints, a push-oriented design that hides the boundary would weaken the very guarantees that make the result scientifically interpretable.

\subsection{Limitations and Scope}

This paper establishes architectural feasibility with production validation; it does not claim that every aspect of virtual-QPU service export is solved. First, the empirical study is centred on a single production Slurm site and a single production simulation path based on Qiskit-Aer/cuQuantum. This is sufficient to demonstrate that the architecture is concrete, but broader multi-site and multi-backend evaluation remains future work. The portability claim should be read as architectural localisation of site-specific logic behind a scheduler adapter, not as a completed empirical survey.

Second, \sys provides authoritative lifecycle semantics rather than a proof of strong exactly-once execution in the full distributed-systems sense. The evaluated claim, heartbeat, and recovery mechanisms preserve task authority and prevent duplicate terminal publication in the tested scenarios. They do not eliminate every ambiguity that can arise from scheduler, filesystem, node, or infrastructure failures. In the target environment, explicit and auditable recovery is part of the correctness model.

Third, the exported device view is snapshot-based rather than continuously synchronised. This choice gives execution hermeticity, temporal clarity, and replayability, but workloads requiring continuous live-state tracking may require a different rebinding policy. Fourth, the deployment assumes a cooperative site boundary in which an unprivileged agent may run on a login node, initiate outbound coordination, submit jobs through the normal scheduler path, and manage local artifacts under site policy.

These limitations point to clear next steps: broader multi-site validation, richer contention studies, direct comparison against manual scheduler workflows, additional scheduler and backend adapters under the same \texttt{SimulatorPort} abstraction, and integration with partitioned or virtualised device abstractions.

\subsection{Conclusion}

We presented \sys, a service-export architecture for exposing HPC-hosted quantum simulation as an interactive virtual-QPU service. \sys resolves the mismatch between device-oriented quantum software and scheduler-mediated HPC by separating service authority from execution realisation, coordinating solely through outbound agents, and binding a topology- and calibration-aware snapshot at claim time as an immutable execution contract.

The broader systems result is that secure, batch-scheduled supercomputers can export semantically rich interactive services without ceasing to operate like HPC systems. The quantum result is that the exported contract can remain faithful to topology, native gates, and calibration state. Together, these results show how HPC-hosted simulation can become part of an interactive, device-oriented quantum software ecosystem and support exploration of both current and future quantum architectures.

\section*{Acknowledgements}
This work was supported by resources provided by the Pawsey Supercomputing Research Centre with funding from the Australian Government and the Government of Western Australia. It was carried out within the Pawsey Supercomputing Research Centre's Quantum Supercomputing Innovation Hub, made possible by a grant from the Australian Government through the National Collaborative Research Infrastructure Strategy (NCRIS). Computational resources were provided by the Pawsey Supercomputing Research Centre's Setonix Supercomputer (\href{https://doi.org/10.48569/18sb-8s43}{https://doi.org/10.48569/18sb-8s43}), with funding from the Australian Government and the Government of Western Australia.

AI-assisted tools were used during the development of portions of the software implementation. All generated code and related documents were reviewed, validated, and modified by the authors.
\section*{Data Availability}
The code and evaluation artefacts are publicly available at \url{https://github.com/PawseySC/HPC-vQPU}.

\clearpage
\section*{Supplementary Material}
\addcontentsline{toc}{section}{Supplementary Material}

\renewcommand{\thesection}{S\arabic{section}}
\renewcommand{\thesubsection}{\thesection.\arabic{subsection}}
\renewcommand{\thesubsubsection}{\thesubsection.\arabic{subsubsection}}
\renewcommand{\thefigure}{S\arabic{figure}}
\renewcommand{\thetable}{S\arabic{table}}
\renewcommand{\theequation}{S\arabic{equation}}
\renewcommand{\theHsection}{S\arabic{section}}
\renewcommand{\theHsubsection}{S\arabic{section}.\arabic{subsection}}
\renewcommand{\theHsubsubsection}{S\arabic{section}.\arabic{subsection}.\arabic{subsubsection}}
\renewcommand{\theHfigure}{S\arabic{figure}}
\renewcommand{\theHtable}{S\arabic{table}}
\renewcommand{\theHequation}{S\arabic{equation}}

\setcounter{section}{0}
\setcounter{subsection}{0}
\setcounter{subsubsection}{0}
\setcounter{figure}{0}
\setcounter{table}{0}
\setcounter{equation}{0}

\newpage
\section{Extended Background and Flow Details}
\label{app:background-details}

The material expands the simulator landscape, the access-model mismatch, device snapshot fields, and execution-plane boundary details.

\subsection{Simulation and Access-Model Context}
\label{app:background-motivation}

Quantum simulation is increasingly asked to do more than reproduce idealised quantum behaviour. It supports algorithm validation, compiler and routing exploration, noise-aware benchmarking, calibration-sensitivity studies, and investigation of future processors by varying topology, native gates, and error parameters~\cite{Lewis2023,IBMQuantum2026,Wang2022,preskill2018quantum,Haener2017,JavadiAbhari2024,bayraktar2023cuquantum}. As simulators adopt multi-GPU and multi-node execution models, the core challenges become memory pressure, communication overhead, accelerator mapping, scheduler interaction, and distributed runtime coordination, not merely local execution.

The evolution of simulator stacks makes this visible. NVIDIA's cuQuantum stack is explicitly designed for accelerated simulation and includes multi-GPU and multi-node deployment paths~\cite{bayraktar2023cuquantum}. Qiskit Aer supports GPU execution, MPI execution, and running with multiple GPUs and nodes~\cite{QAD2026,JavadiAbhari2024}. SV-Sim frames state-vector simulation as a PGAS-based HPC problem targeting single-node and multi-node CPU/GPU clusters, with emphasis on irregular communication costs~\cite{Li2021,Lykov2021}. These systems demonstrate that serious quantum simulation is already an HPC workload; \sys addresses how to make that workload available through a device-aware service boundary.

Traditional HPC and commercial cloud models remain incomplete for this purpose. HPC systems such as LUMI and Setonix expose accounts, filesystems, batch scripts, modules, queues, and allocation policies~\cite{LUMIConsortium2026,PSRC2023}. Cloud services expose simpler APIs, but communication-heavy simulation reintroduces explicit placement, storage, and network configuration concerns, such as low-latency instance placement and separately provisioned parallel storage. For example, cloud HPC deployments often depend on mechanisms such as Elastic Fabric Adapter, placement-aware instance selection, and FSx for Lustre rather than on a universally integrated supercomputing substrate~\cite{AWS2026a,AWS2026b}. The required model is therefore a hybrid: cloud-like device semantics at the boundary and scheduler-mediated HPC execution underneath.

\subsection{Device Model Field Semantics}
\label{app:device-model-details}

The snapshot in Eq.~\eqref{main-eq:snapshot} is intentionally graph-structured. $Q=\{q_0,\dots,q_{n-1}\}$ is the exposed qubit set. $G=(Q,E)$ is the coupling graph, with directed edges when gate availability or calibration differs by direction. $\mathcal{G}_{\mathrm{native}}$ is the set of operations treated as directly executable on the target device. $\Theta_Q$ stores per-qubit execution parameters, including coherence times, single-qubit error, readout error, and operational status. $\Theta_E$ stores per-edge parameters, including two-qubit gate type and error rate.

This representation is not a full pulse-level hardware description. It contains the information needed to preserve virtual-device execution semantics: connectivity, operation constraints, and calibration-derived noise behaviour. It also supports heterogeneous and hypothetical architectures without topology-specific code. Heavy-hex devices, grids, rings, stars, and synthetic designs can all be represented as snapshots consumed by the same validation and execution paths.

The time index is essential. The meaning of ``run on device $d$'' is underspecified unless the relevant device state is pinned to a particular moment or version. If a scheduled job received only a circuit and device identifier, it would need a later live lookup to recover device meaning, creating a fragile dependency on external reachability and mutable state. If device state were ignored, the system would cease to be device-aware. The snapshot is the minimal portable contract that avoids both failures.

\section{Design Boundary Notes}
\label{app:design-details}
This appendix records boundary conditions that support the main argument but would unnecessarily interrupt the flow of the body text. Its organising principle is the same as in the paper as a whole: design states architectural requirements, implementation selects concrete realisations that satisfy them, and evaluation tests the consequences of those choices. The role of the appendix is therefore not to introduce new claims but to close interpretive gaps, expose secondary mechanisms, and keep the argument auditable.

\subsection{Claim Atomicity and Store Realisation}
\label{app:design-claim-store}
Invariant~D3 states an architectural requirement rather than prescribing a specific storage engine. In the current prototype, the indivisibility of ownership binding and state transition is realised through store-side write serialisation. A scaled deployment would realise the same invariant through transactional row-level locking or an equivalent store primitive. The architectural point is therefore not SQLite versus PostgreSQL, but that ownership and lifecycle transition share the same linearisation point.

\subsection{Snapshot Freshness Policy}
\label{app:design-cache}

The TTL-governed cache is keyed by device identifier because device state is read far more frequently than it is mutated. Administrative mutations, including calibration refresh, topology change, and qubit state update, which invalidate the corresponding cached entry immediately upon commit, so post-mutation reads are never masked by the cache. Read-only workloads during periods of device stability therefore pay the reconstruction cost at most once per TTL window, while the authoritative snapshot used for execution is always bound from the current authoritative state at claim time.

\subsection{Observability Projection Mechanism}
\label{app:design-sse}

The design requirement behind D9 is not merely that events exist, but that they be exported without turning observation into a second control path. In the implementation, the control plane realises this through Server-Sent Events with bounded-memory fan-out, keep-alive heartbeats, slow-consumer eviction, and replay support for reconnecting clients. These choices matter operationally, but at the design level, they are subordinate to the stronger property already stated in the main text: the observation path is derived entirely from committed control-plane state and does not introduce any communication into the HPC environment.

\section{Additional Implementation Details}
\label{app:impl-details}
This appendix gathers implementation material that supports the main implementation narrative but no longer needs to remain in the body. The material is organised from the service-layer boundary, to execution semantics, to failure and recovery, and finally to an implementation-to-invariant synthesis. This ordering is intentional: it preserves the same explanatory logic as Section~\ref{main-sec:impl} and prevents the appendix from reading like an unordered repository of residual detail.

\subsection{Service-Layer Contract Families}
\label{app:impl-contracts}

The port layer can be understood as three contract families. The first governs lifecycle and identity: \emph{TaskStorePort} exposes task-state transitions without revealing storage semantics, \emph{EventStorePort} provides append-only event persistence and time-range queries, and \emph{AuthPort} validates credentials and returns a typed principal that carries user and role information. The second governs device-state access: \emph{DeviceStorePort} exposes both authoritative device state and a cached snapshot view, while \emph{CachePort} provides TTL-keyed get/set/invalidate operations. The third governs execution: \emph{SimulatorPort} is the sole interface through which the control plane can request circuit evaluation. The current implementation instantiates these contracts with SQLite-backed stores, an in-memory cache, a JSONL append-only event log, and API-key authentication, but the service layer is written entirely against the contracts rather than against those concrete choices.

This organisation is load-bearing for correctness, not just for maintainability. Because the service layer depends only on abstract ports, its behaviour cannot vary as a function of which simulator implementation is installed, which storage backend persists task state, or which cache implementation serves device metadata. In particular, the control plane has no dependency path by which scheduler state or backend-specific runtime details can enter the task lifecycle logic. The resulting isolation is therefore visible in the dependency structure itself: concrete adapters are confined to the composition root, while the domain core is parameterised only by port contracts.

\subsection{Execution-Plane Operational Properties}
\label{app:impl-agent-properties}

The execution-plane agent is deliberately narrower than a general worker service. Each claimed task is materialised into its own run directory with its own bound snapshot, scheduler-backed execution unit, and result artifact, so failures remain task-local rather than contaminating other executions. The agent is also stateless with respect to authoritative task meaning: it owns no durable task database and reconstructs its working view from the control plane, scheduler evidence, and local artifacts after restart. Site-specific launch choices, including scheduler adapter, walltime, account strings, module stack, and environment activation, remain local configuration. Conversely, the agent is blind to service-level concerns such as device management, authentication policy, public REST resources, and event projection. Its reusable interface is the narrow outbound protocol for claim, heartbeat, and report.

\subsection{Submission-Time Admissibility Checking}
\label{app:impl-viability}
Submission-time admissibility checking exists to answer a specific control-plane question: can a request be rejected immediately, before any scheduler resource is committed, based on the currently advertised device description? This is the implementation point at which Invariant~D6 becomes operational. The check must therefore satisfy two requirements simultaneously: it must be fast enough to remain on the interactive API path, and it must be expressed purely in terms of the device abstraction rather than topology-specific logic.

Concretely, the admissibility test evaluates three structural conditions against the currently advertised device description for the target device. First, every gate symbol appearing in the submitted circuit must belong to the native gate set $\mathcal{G}_{\mathrm{native}}$, excluding syntactic constructs such as barriers, measurements, and resets that do not encode backend-specific gate semantics. Second, every referenced qubit index must map to a qubit whose state is \textsc{Online}. This preserves the distinction between physical presence and operational availability: a qubit may still exist in the coupling graph yet remain inadmissible because the virtualised device marks it as unavailable. Third, every multi-qubit operation must respect the edge set $E$ of the advertised coupling graph, including directionality where the underlying interaction model is asymmetric. A two-qubit gate is therefore admissible only if its ordered operand pair corresponds to an \textsc{Available} coupling in the graph.

The important implementation choice is that this check is intentionally \emph{syntactic} and \emph{linear-time}. Rather than importing the full compiler stack into the submission path, the service performs a lightweight lexical pass over the submitted source to extract gate symbols and referenced qubit indices, and then evaluates the three predicates directly against the currently advertised device description. This keeps the hot-path cost proportional to the circuit description size, avoids entangling admission logic with simulator or transpiler dependencies, and preserves the architectural separation described in Section~\ref{main-sec:impl-layers}. More importantly, it ensures that admissibility remains topology-parametric by construction: the validator consumes only $\mathcal{G}_{\mathrm{native}}$, qubit states, and the coupling graph, regardless of whether the advertised device describes a heavy-hex processor, a grid, a star, or a hypothetical architecture assembled for exploratory studies.

This design deliberately separates \emph{early rejection} from \emph{final execution binding}. Submission-time admissibility is evaluated against the currently advertised device description to reject obviously impossible requests without incurring scheduler delay, but it does not establish the immutable execution contract. The authoritative execution snapshot is still bound at claim time, in accordance with Invariant~D5. This distinction matters because the read path may be served from a bounded-staleness cache, whereas the execution payload must reflect authoritative state at the moment ownership is transferred. Submission-time checking improves responsiveness and avoids wasted scheduling effort, while claim-time binding preserves semantic correctness as devices evolve.

The same validation logic can be exposed as a viability-check endpoint for clients and automated workflows that need pre-flight feedback without creating a durable task. This preserves the service-level distinction between asking whether a circuit is currently compatible with a virtual device and committing work into the scheduler-backed lifecycle.

The result is a deliberately two-stage validation pipeline. The first stage is a lightweight, graph-syntactic filter at the API boundary that eliminates requests incompatible with the currently visible device abstraction. The second stage occurs inside the hermetic compute-node runner: after parsing and transpilation against the bound execution snapshot, the backend performs its normal pre-flight validation on the executable circuit before simulation begins. This latter stage can detect violations that are not visible in the submitted syntax alone, such as incompatibilities introduced by decomposition, routing, or backend-specific compilation. The two stages serve different purposes: the first minimises latency on the control path, while the second closes semantic completeness on the execution path. Together, they realise D6 without sacrificing either interactivity or correctness.

\subsection{Snapshot-Derived Noise Model}
\label{app:impl-noise}
For \sys, a device snapshot is not merely a topology descriptor; it is the complete simulation contract for a virtual QPU at a particular time. A simulator that ignores the calibration fields carried by a bound snapshot can reproduce connectivity, but not device behaviour, thereby reducing virtualisation to an idealised functional model. To preserve the operational meaning of a bound execution contract, \sys derives the noise model directly from the calibration parameters carried by the bound snapshot. The resulting execution semantics are therefore snapshot-specific rather than backend-global: two tasks with identical circuits but different bound snapshots may legitimately produce different output distributions because they represent different virtualised device states.

Concretely, the derived noise model $\mathcal{N}(\delta_i)$ is assembled from three calibration-driven channel families. First, each qubit $q_i \in Q$ with single-qubit error parameter $\epsilon_{\mathrm{1q}}(q_i) > 0$ receives a depolarising channel on each \emph{physical} single-qubit gate acting on that qubit. Gates realised that virtual frame updates or timing directives, such as $R_z$, identity, or delay, are excluded because they do not correspond to a physical control pulse and should not accumulate pulse-level gate error in the simulator. Second, each qubit with readout assignment error $r(q_i) > 0$ receives a symmetric readout channel
\[
\begin{bmatrix}
1-r(q_i) & r(q_i) \\
r(q_i) & 1-r(q_i)
\end{bmatrix},
\]
with $r(q_i)$ clamped to $[0,0.5]$ to preserve a valid stochastic operator. Third, each directed coupling $(q_i,q_j)\in E$ with edge error rate $\epsilon_{ij} > 0$ receives a two-qubit depolarising channel attached to the gate type $g_{ij}$ declared for that edge. Directionality is preserved throughout this construction: $\epsilon_{ij}$ and $\epsilon_{ji}$ are treated as distinct calibration quantities, allowing the virtual device to reflect asymmetric interaction quality where the underlying hardware exhibits directional coupling behaviour.

This construction is intentionally snapshot-total. If the bound snapshot carries no usable calibration fields, the derived model collapses to $\mathcal{N}(\delta_i)=\varnothing$, and execution proceeds ideally. This is not a special execution mode selected by separate configuration; it is the degenerate case of the same derivation pipeline. As a result, ideal, partially calibrated, and fully noise-annotated simulations are all expressed through the same abstraction. The control plane does not switch between different simulator semantics; it always evaluates against the noise semantics implied by the bound snapshot, whether that implication is empty or rich.

The same derivation logic is executed in both runtime paths: the control-plane simulator and the hermetic compute-node runner construct $\mathcal{N}(\delta_i)$ from the same snapshot fields using the same exclusion rules, parameterisation, and clamping policy. This symmetry is not an implementation convenience but part of the invariant argument. If the two paths were allowed to derive different noise models from the same bound snapshot, then the meaning of the execution contract would become path-dependent, violating Invariant~D5 even when the bound snapshot itself was identical. Noise derivation is therefore treated as part of contract materialisation rather than as backend-local post-processing.

A broader consequence is that snapshots become first-class historical records rather than transient cache entries. Because execution binds a concrete snapshot and derives noise from that bound record, \sys naturally supports a data-warehouse-style view of device management: snapshots can represent different processors, different calibration epochs of the same processor, or deliberately constructed hypothetical devices, all while being consumed through one uniform execution interface. A user may replay an experiment against an older snapshot to reproduce prior behaviour, compare two calibration epochs of the same hardware under identical workloads, or evaluate an as-yet-unbuilt architecture by registering a synthetic topology and calibration profile. In all three cases, the simulator is driven not by mutable live backend state but by an explicit, versionable device record.

This historical binding is what makes experimental results auditable and repeatable. A task result need not be interpreted against whatever the vendor backend looks like today; it can always be traced back to the exact snapshot from which both admissibility and noise were derived. In practice, this gives \sys two capabilities that are difficult to obtain from direct access to live QPUs alone: longitudinal analysis over changing calibration regimes, and exact replay of prior virtual experiments after the physical hardware has drifted or been reconfigured. The virtual QPU is therefore not just a convenient access abstraction; it is a persistent execution view over a versioned corpus of device states.

\subsection{Cross-Domain Fault Provenance and Error Projection}
\label{app:impl-errors}

The remaining implementation details concern the failure boundary. The point is not simply that errors are reported, but that heterogeneous evidence from different operational domains is normalised without weakening the control plane's authority over lifecycle state.

For each task $\tau_i$, the control plane maintains a single authoritative lifecycle state $\sigma(\tau_i)$. Crossing the HPC boundary does not change this authority model, but it does enlarge the set of failure origins that may influence how $\sigma(\tau_i)$ evolves. In \sys, these execution-side observations are treated as evidence $e_i$, not as state: admission failures, scheduler anomalies, runner exceptions, missing or malformed result artifacts, and agent-liveness loss arise in different domains and at different times, but none of them is, by itself, an authoritative task outcome. The implementation problem is therefore not to enumerate error cases exhaustively, but to preserve \emph{fault provenance} while translating heterogeneous evidence into exactly one legal control-plane action.

The boundary stages this translation. Failures detected before ownership is established remain entirely within the control plane. Invalid credentials, illegal lifecycle transitions, and circuits that violate the submission-time admissibility predicate are rejected synchronously and do not enter the execution domain at all. This early containment is semantically important: user-correctable requests and protocol-illegal operations should not be allowed to degrade into scheduler-visible failures, because doing so would conflate invalid work with failed work.

After ownership has been established, fault handling becomes cross-domain. The scheduler may indicate that the submitted job never started, was cancelled, or terminated abnormally; the hermetic runner may raise an exception, exhaust memory, or fail to emit a valid result artifact; and the agent may disappear while owned work remains in flight. These are all execution-side observations contributing to $e_i$. Their role is evidentiary only. The agent reconciles them and attempts one legal lifecycle action through the authoritative transition boundary: publish completion, report typed failure, or leave the task eligible for explicit recovery if ownership has lapsed. In this way, heterogeneous failures remain local where they arise, yet still become service-visible through a single, controlled projection path.

The crucial requirement is that this projection preserve \emph{recovery semantics}, not merely human-readable diagnostics. Different origins permit different correct responses: an admission failure requires a changed request; a scheduler-side infrastructure failure may justify requeueing; a backend runtime failure may call for reduced circuit size or larger resource allocation; an agent-liveness failure may require explicit operator-driven recovery. If all such cases were collapsed into a single undifferentiated failure outcome, the system would destroy the information needed by both users and the orchestration logic to choose the correct next action. \sys therefore projects failures through a uniform wire schema that carries a machine-readable code, message, optional structured detail, a correlation identifier, and a timestamp. This is not presentation polish layered over exceptions; it is the mechanism by which cross-domain provenance survives normalisation into a coherent service-visible error surface.

\subsection{Coherence Under Combined Failures}
\label{app:impl-coherence}

The harder systems question is not how any single fault is classified, but whether the task automaton remains coherent when multiple faults occur simultaneously. In \sys, coherence means that temporary disagreement across domains may occur, but the control plane still converges to a single valid $\sigma(\tau_i)$ without violating ownership or lifecycle legality. This property matters because execution spans multiple partially independent substrates: the task store, the scheduler, the local run-directory filesystem, and the agent process may fail or recover on different timelines.

A representative case is as follows: an agent claims $\tau_i$, submits the corresponding Slurm job via Dask, reports $\Running$, and then crashes before the job completes. The job may nonetheless run to completion and produce a valid result artifact. From the control plane's perspective, however, the only authoritative fact is that $\sigma(\tau_i)=\Running$ and the heartbeat has become stale. Because Invariant~D1 prohibits direct control-plane reachability into the cluster, the service cannot independently determine whether the job is still executing, has already finished, or has failed silently. The system therefore resolves such cases by \emph{bounded explicit recovery} rather than by pretending to infer hidden remote truth. Once ownership liveness expires, the task becomes eligible for policy-driven or operator-driven intervention, such as \textsc{Requeue} or \textsc{Force-Fail}.

If the agent later returns, recovery is not handled by a separate crash-specific subsystem. The restarted agent simply re-enters the normal reconciliation loop: it re-observes scheduler state, inspects local run directories, and compares those observations against the still-authoritative control-plane state. If it finds that a completed artifact exists for a task still recorded as $\Running$, it attempts the same standard finalisation path used during normal operation. Recovery is therefore a consequence of the ordinary reconciliation protocol, not a bolt-on exception path.

This is what keeps the combined-failure state space tractable. The relevant domains fail independently, but they do not overwrite one another's truth. A scheduler anomaly cannot directly mutate the task store; a task-store transition cannot fabricate a result artifact; an agent crash does not alter the state of an in-flight Slurm job. Each domain contributes evidence, while the task automaton remains the sole arbiter of authoritative state. As a result, disagreement may be temporary, but it is structurally bounded: each reconciliation step either commits one legal transition of $\sigma(\tau_i)$ or discovers that no further action is currently justified. The system may observe partial and stale evidence across domains, yet it still converges to a single coherent lifecycle outcome because only the control plane is allowed to decide what that outcome is.

\subsection{Implementation-to-Invariant Synthesis}
\label{app:impl-map}
The preceding implementation supplements elaborate specific mechanisms. Table~\ref{tab:impl-map} closes the appendix by showing how those mechanisms align with the invariants introduced in Section~\ref{main-sec:design}. The table is meant as a synthesis aid rather than a replacement for the section-by-section argument.

\begin{table*}[t]
\centering
\caption{Mapping of design invariants to implementation mechanisms.}
\label{tab:impl-map}
\small
\begin{tabular}{lll}
\toprule
Invariant & Mechanism & Section \\
\midrule
D1 (outbound-only) & Capability-partitioned two-artifact deployment & \ref{main-sec:impl-isolation} \\
D2 (authoritative DFA) & Serialised authoritative task store with static transition relation & \ref{main-sec:impl-claim} \\
D3 (exclusive ownership) & Atomic \textsc{Claim} as ownership linearisation point & \ref{main-sec:impl-claim} \\
D4 (bounded liveness) & Heartbeat expiry plus reconciliation-driven recovery & \ref{main-sec:impl-agent} \\
D5 (immutable contract) & Snapshot-parameterised execution contract (\texttt{SimulatorPort} + bound snapshot replay) & \ref{main-sec:impl-layers} \\
D6 (topology-parametric) & Graph-parametric admissibility over snapshot topology & \ref{app:impl-viability} \\
D7 (TTL-governed cache) & Dual-view device state with invalidate-on-mutate caching & \ref{main-sec:impl-layers} \\
D8 (hermetic runner) & Artifact-local functional execution boundary with import isolation & \ref{main-sec:impl-runner} \\
D9 (service-side observability) & Authoritative event projection with bounded-memory SSE fan-out & \ref{main-sec:impl-sse} \\
\bottomrule
\end{tabular}
\end{table*}

Table~\ref{tab:impl-map} summarises the \emph{primary} mapping from design invariants to implementation mechanisms. The mapping is not intended to be a one-to-one index of subsections: several subsections refine, compose, or stress-test earlier mechanisms rather than introduce new invariant-enforcement boundaries. In particular, Appendix~\ref{app:impl-viability} and Appendix~\ref{app:impl-noise} elaborate how the bound snapshot is used to realise D5--D6 during execution; Section~\ref{main-sec:impl-slurm} localises site-specific scheduler variability behind a narrow integration boundary; and Appendix~\ref{app:impl-errors}--\ref{app:impl-coherence} show how the same control-plane invariants remain coherent under cross-domain failure.

A central property of the implementation is that these mechanisms are decoupled. Task-store atomicity (D2--D3), reconciliation-driven liveness (D4), snapshot-bound execution semantics (D5--D7), hermetic execution (D8), and service-side event projection (D9) are enforced by distinct boundaries with non-overlapping responsibilities. This makes the correctness argument compositional and the failure surface local: breaking one mechanism does not silently violate an invariant enforced by another. In this respect, \sys is not an ad hoc combination of a web server and a batch script, but a deliberately structured realisation of the design invariants introduced earlier.

\section{Additional Evaluation Details}
\label{app:eval-method}
This appendix reorganises the supplementary evaluation material into a continuous architecture-facing narrative. It begins with experimental context and the relationship between deployment experiments and the regression suite, then records the fuller numerical and control evidence underlying the main claims, and finally closes with operational supplements and an evaluation-to-invariant synthesis table. The purpose is not to add a second evaluation story, but to preserve inspectability without overloading the main text.

\subsection{Experimental Context and Layered Evidence}
\label{app:eval-context}

The evaluation appendix first establishes how the experiments were conducted and how they should be interpreted relative to the regression suite.

\subsubsection{Experimental Context}
\label{app:eval-setup}
All end-to-end experiments are conducted on the deployed \sys instance at the Pawsey Supercomputing Research Centre. The cloud-visible control plane (\sysserver) is hosted on Nectar infrastructure operated by Pawsey, while the execution plane (\sysagent) runs on a Setonix login node and submits jobs through Slurm to the Setonix Quantum partition. Unless otherwise stated, execution uses backend \texttt{qiskit-ella}, which invokes Qiskit-Aer with cuQuantum on Setonix GPUs. The agent polls the control plane at approximately 30-second intervals.

The primary evaluation target is \texttt{ibm-fez-0309}, a 156-qubit virtual device instantiated from IBM Fez calibration data captured on 2026-03-09. The snapshot encodes the heavy-hex topology, CZ as the native two-qubit gate, per-qubit readout error rates, and per-edge two-qubit gate error rates. When a noise-free reference is required, we use \texttt{ibm-fez-ideal}, which preserves the same 156-qubit topology and native gate set while setting all error fields to null. This paired configuration isolates the effect of calibration semantics without changing the validation or execution path.

To avoid confounding from backend-dependent compilation choices, all evaluation circuits are expressed directly in the native gate set $(\mathrm{CZ}, \mathrm{SX}, \mathrm{RZ})$. Differences in latency or output distributions can therefore be attributed to the architectural mechanisms under study rather than to stochastic transpilation behaviour. Measurements are collected at two instrumentation points. The control plane records admission and lifecycle-event timestamps, while the runner records parse time, noise model construction time, transpilation time, and simulation time. Together, these measurements allow us to separate service-layer cost from scheduler delay and GPU execution cost across the experiments that follow.

\subsubsection{Relationship to the Regression Suite}
\label{app:eval-correctness}

The evaluation evidence in this paper is intentionally stratified. The end-to-end experiments in Section~\ref{main-sec:eval} validate the architectural claims under realistic deployment conditions, but they are not designed to enumerate all forbidden transitions or race windows exhaustively. That role belongs to the regression suite. The implementation is additionally validated by 258 deterministic tests spanning exclusive claim under concurrent polling, rejection of illegal state transitions, terminal-state absorption, duplicate-completion rejection, TTL-governed cache invalidation with immediate post-mutation consistency, SSE ordering and replay, bounded-memory fan-out, keep-alive delivery, and topology-parametric validation across multiple graph families.

This division of labour is itself an important methodological point. The regression suite provides \emph{closure over the design state space}: it verifies that invalid transitions are rejected, that terminal states absorb future events, that cache mutations become visible at the intended boundary, and that observability protocols behave deterministically under controlled conditions. The production experiments provide something different: they test whether these same invariants remain meaningful when embedded in a real operational context involving queue delays, GPU execution, calibration mutation, and agent failures. The former is structural evidence; the latter is semantic evidence.

Seen this way, the two layers are not redundant. A pure regression story would show that the mechanisms are internally consistent, but not that they survive real scheduler-mediated deployment. A pure deployment story would show interesting behaviour but leave open whether that behaviour is accidental, under-constrained, or incompletely covered. Their combination is what makes the evaluation strong. The regression suite gives the operational experiments interpretive discipline, while the deployment experiments show that the regression-level invariants continue to govern system behaviour when the environment becomes asynchronous, delayed, and partially observable.

This layered structure also explains why the paper's strongest claims are architectural rather than benchmark-centric. Bounded service overhead, claim-time freshness, calibration-aware execution, and explicit crash recovery are not independent anecdotes; they are manifestations of the same invariant-driven design viewed under different evidentiary lenses. The regression suite proves that the control logic has the right shape. The deployment experiments prove that the same shape survives contact with the real HPC boundary. Together, they amount to a stronger argument than either form of evidence could provide alone.

\subsection{Additional Evidence for Bounded Service Overhead}
\label{app:eval-latency}

The main text presents the decisive bounded-overhead claim. This subsection records the fuller numerical decomposition and its interpretation so that the latency argument remains inspectable without expanding the main evaluation narrative.

\subsubsection{Tabulated Latency Breakdown}
\label{app:eval-latency-tables}

Tables~\ref{tab:latency} and~\ref{tab:sim-breakdown} provide the detailed numerical decomposition underlying the bounded-overhead argument in Section~\ref{main-sec:eval}. The first table separates end-to-end service-path time into admission, queue-to-claim delay, execution, and poll-visible completion. The second table decomposes the execution phase itself into parse, noise-model construction, transpilation, and GPU simulation, with the residual capturing the remaining claim-to-completion time outside the explicitly instrumented runner stages.

\begin{table}[!htp]
  \centering
  \caption{End-to-end latency decomposition (seconds, mean of 5 trials). Interface overhead is defined as $T_\text{admit}+T_\text{poll}$, and OH\% denotes its fraction of end-to-end time.}
  \label{tab:latency}
  \small
  \setlength{\tabcolsep}{3.5pt}
  \begin{tabular}{@{}rrrrrrrr@{}}
    \toprule
    $n$ & $T_\text{e2e}$ & $T_\text{adm}$ & $T_\text{q-clm}$ & $T_\text{exec}$ & $T_\text{poll}$ & OH & OH\% \\
    \midrule
    28 & 54.82 & 0.42 & 21.97 & 30.30 & 2.79 & 3.21 & 5.9\% \\
    29 & 58.77 & 0.39 & 26.02 & 30.37 & 2.65 & 3.04 & 5.2\% \\
    30 & 58.16 & 0.38 & 26.53 & 30.30 & 1.59 & 1.97 & 3.4\% \\
    31 & 58.91 & 0.42 & 25.84 & 30.31 & 2.98 & 3.40 & 5.8\% \\
    32 & 58.97 & 0.42 & 27.07 & 30.31 & 1.80 & 2.21 & 3.7\% \\
    \bottomrule
  \end{tabular}
\end{table}

\begin{table}[!htp]
  \centering
  \caption{Execution-phase breakdown within each claimed job (seconds, mean of 5 trials). Here $t_\text{total}=t_\text{parse}+t_\text{noise}+t_\text{xpile}+t_\text{sim}$, and Residual $= T_\text{exec}-t_\text{total}$. The residual, therefore, captures uninstrumented claim-to-completion time outside the explicitly measured runner stages.}
  \label{tab:sim-breakdown}
  \small
  \begin{tabular}{rrrrrrrr}
    \toprule
    $n$ & $t_\text{parse}$ & $t_\text{noise}$ & $t_\text{xpile}$ & $t_\text{sim}$ & $t_\text{total}$ & Residual & $t_\text{sim}$ ratio \\
    \midrule
    28 & 0.65 & 0.61 & 0.14 & 1.32 & 2.72 & 27.58 & --- \\
    29 & 0.60 & 0.58 & 0.13 & 2.76 & 4.07 & 26.29 & 2.09$\times$ \\
    30 & 0.60 & 0.58 & 0.13 & 5.30 & 6.62 & 23.68 & 1.92$\times$ \\
    31 & 0.61 & 0.58 & 0.14 & 10.94 & 12.27 & 18.05 & 2.06$\times$ \\
    32 & 0.61 & 0.58 & 0.14 & 23.77 & 25.11 & 5.20 & 2.17$\times$ \\
    \bottomrule
  \end{tabular}
\end{table}

\subsubsection{Interpretation of the Bounded-Overhead Regime}
\label{app:eval-latency-interpretation}
The strongest conclusion from Tables~\ref{tab:latency} and~\ref{tab:sim-breakdown} is not merely that the observed overhead is bounded, but that its growth behaviour is cleanly separated from the growth behaviour of the simulation itself. This is the key system's result. As qubit count increases, the workload-dependent cost remains confined to the compute boundary, while the service-export costs remain localised to admission, dispatch visibility, and completion visibility. In particular, $T_\text{admit}$ stays near 0.4\,s across all configurations, and $T_\text{poll}$ remains bounded by the client polling interval. Neither term shows any growth trend as $n$ increases. The architecture adds latency without introducing a second workload-dependent scaling law.

The nearly constant end-to-end latency over $n=28$--$32$ must therefore be interpreted carefully. It is not evidence that the underlying computation is constant, but evidence that this experiment operates in a fixed-cost-dominated regime. The runner-side breakdown makes this explicit. Inside the claimed job, parse, noise-model construction, and transpilation remain essentially constant at about 1.3\,s total, whereas GPU simulation grows from 1.32\,s to 23.77\,s, with per-qubit growth factors of 2.09, 1.92, 2.06, and 2.17, closely matching the expected $\Theta(2^n)$ trend. At the same time, the residual non-simulation portion of $T_\text{exec}$ falls from 27.58\,s at $n=28$ to 5.20\,s at $n=32$. That contraction matters because it shows that the fixed cost is additive rather than multiplicative. If the service layer or launch path were introducing a scaling penalty, this residual would grow with $n$ or at least remain proportionally dominant. Instead, it is rapidly amortised as simulation cost rises.

This reveals a two-regime structure. In the first regime, which is the one observed directly in Table~\ref{tab:latency}, bounded dispatch and startup costs dominate, so $T_\text{e2e}$ appears almost flat even while the underlying GPU work is growing. In the second regime, once $t_\text{sim}$ exceeds that bounded floor, end-to-end latency must begin to track the exponential simulation curve directly. The present data already show the transition toward that regime: by $n=32$, the simulation term has grown to 23.77\,s while the residual has shrunk to 5.20\,s, so the compute component is close to overtaking the constant region. The important point is that \sys does not suppress scaling; it exposes the expected scaling once a bounded deployment-level floor has been amortised.

The dominant constant region should also be interpreted carefully. It should not be attributed solely to Slurm, because the queue-to-claim interval itself already lies in the 22--27\,s range and is shaped primarily by the agent's 30\,s polling cadence. The observed floor is therefore better understood as the chosen dispatch granularity of the current polling-based deployment, combined with normal execution-path startup costs. This is an operational trade-off consistent with the architecture's outbound-only, scheduler-compatible design: the system avoids inbound listeners, resident brokers, and tighter coupling to the HPC environment at the cost of coarser dispatch cadence. The constant floor is therefore deployment-contingent rather than architecture-intrinsic.

A more revealing measure than overhead as a percentage of end-to-end time is the ratio between service-interface overhead and useful GPU work. Using $T_\text{admit}+T_\text{poll}$ as the interface-visible overhead, that ratio drops from approximately $2.4\times$ at $n=28$ to about $0.09\times$ at $n=32$. This trend matters more than the raw percentages in Table~\ref{tab:latency}: it shows that the exported-service envelope becomes asymptotically negligible relative to the computation it mediates. The architectural claim supported by this experiment is therefore stronger than ``bounded overhead'': \sys preserves the scaling behaviour of the underlying HPC simulator by localising additional cost to bounded, deployment-level stages rather than allowing it to couple to workload growth.

Finally, the decomposition itself depends on the correctness of the authoritative event stream. All intervals except client-side admission are reconstructed from \textsc{TASK\_QUEUED}, \textsc{TASK\_RUNNING}, and \textsc{TASK\_COMPLETED} timestamps. The internal consistency of these measurements across repeated trials also provides indirect evidence for Invariant~D9: the observability layer records lifecycle transitions faithfully enough to support post hoc performance analysis without coupling clients to the HPC execution path.

\begin{figure}[!htp]
  \centering
  \includegraphics[width=\columnwidth]{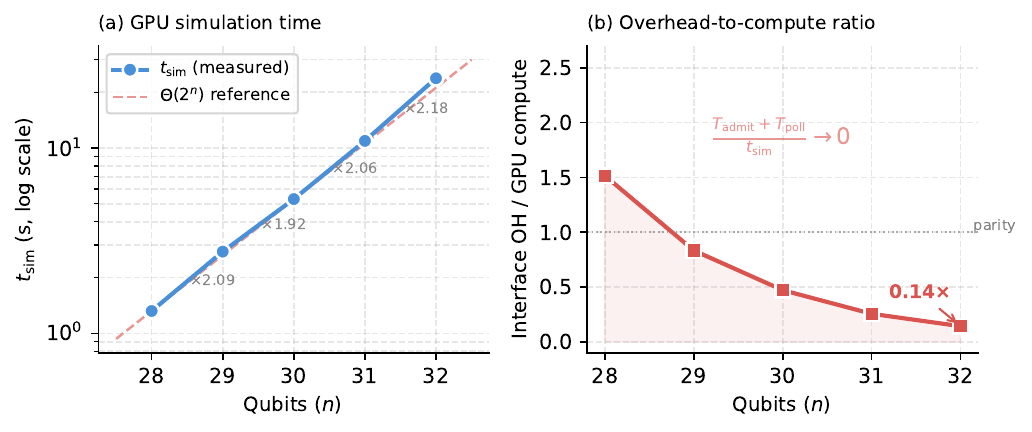}
  \caption{(a)~GPU simulation time on a logarithmic scale with per-qubit
  growth factors annotated; the data closely track the $\Theta(2^n)$
  reference.
  (b)~Interface overhead $(T_\mathrm{admit}+T_\mathrm{poll})$
  normalised by GPU compute cost; the ratio falls below parity by $n{=}29$
  and approaches zero as simulation cost dominates, confirming that the
  service-export envelope becomes asymptotically negligible relative
  to the computation it mediates.}
  \label{fig:scaling-ratio}
\end{figure}

\subsection{Additional Controls for Snapshot Semantics and Binding}
\label{app:eval-snapshot}

The main text keeps only the decisive evidence for invariants D5--D7. This subsection restores the controls and interpretive elaborations that support that claim without burdening the primary evaluation path.

\subsubsection{Additional Discussion of Calibration-Induced Structure}

This is stronger than a generic noisy-versus-ideal comparison across unrelated backends. Because topology and execution path are held fixed, the result rules out alternative explanations in which the output changes merely because a different device family, routing choice, or compiler path was exercised. Instead, the change is induced solely by calibration semantics. In this sense, \sys does not merely transport device descriptors across the HPC boundary; it transports execution-relevant device state whose effects survive delayed, scheduler-mediated execution and become observable at the compute node.

The structure of the noisy output is also informative. If the snapshot acted only as a coarse global fidelity penalty, one would expect a less distinctive redistribution of probability mass. Instead, the leakage pattern is asymmetric and reproducible across trials, which is exactly the behaviour expected if the runner is instantiating a fine-grained contract derived from per-qubit and per-edge calibration fields. The experiment, therefore, validates not only that the snapshot matters, but that it matters at the right semantic resolution.

Finally, this experiment closes the loop on Invariant~D6. Earlier sections argue that admissibility and device interpretation must remain graph-parametric rather than vendor- or topology-specific. Here, the same graph-driven validation, snapshot parsing, and noise-model construction paths operate correctly on a real 156-qubit Fez heavy-hex snapshot without topology-specific branching. The result is therefore not only a fidelity check, but also production-scale evidence that the topology-parametric abstraction remains intact when the snapshot moves from toy graphs to realistic device structures.

\subsubsection{Scene~A: Cross-Device Misbinding Control}
\label{app:binding-scene-a}

\paragraph*{Protocol.}
We use the same amplified two-qubit identity workload as in Section~\ref{main-sec:eval-noise}, which provides a stable output fingerprint under both noisy and ideal snapshots. Scene~A is a mixed-batch control. We alternately submit 8 tasks to \texttt{ibm-fez-0309} and 8 tasks to \texttt{ibm-fez-ideal}, yielding 16 interleaved requests. Because the two devices have identical topology but different calibration content, any accidental cross-binding at claim time would be visible in the output distribution.

\paragraph*{Results.}
Scene~A serves only as a control. Tasks targeting \texttt{ibm-fez-0309} reproduce the same noisy fingerprint established in Section~\ref{main-sec:eval-noise}, with mean TV $= 0.03528$, while tasks targeting \texttt{ibm-fez-ideal} remain exact with TV $= 0$. The noisy mean in Scene~A is effectively identical to the baseline noisy mean of 0.03537 reported in Section~\ref{main-sec:eval-noise}, indicating that queue interleaving does not perturb device identity or cross-bind tasks to the wrong snapshot.

\paragraph*{Interpretation.}
Scene~A establishes identity preservation under concurrency: even when requests are interleaved in the queue, noisy and ideal tasks retain their respective fingerprints. This rules out a simple cross-device misbinding failure mode, but it does not yet identify the moment in the lifecycle when the snapshot becomes fixed.

\subsubsection{Additional Notes on Claim-Time Binding}
The main text focuses on Scene~B because it is decisive, but Scene~A and Scene~B establish different properties. Scene~A is a control for identity preservation under queue interleaving, whereas Scene~B identifies the lifecycle point at which the execution snapshot becomes fixed.

Taken together, the two scenes also strengthen the argument for implementation. Earlier sections separate submission-time admissibility from final execution binding and require that execution-time noise semantics be derived from the bound snapshot rather than from backend-global state. Scene~A shows that interleaving does not induce cross-device misbinding; Scene~B shows that an administrative mutation that occurs before claim changes, which snapshot is subsequently bound. The combined result therefore validates not only claim-time binding in the abstract, but also the end-to-end alignment of the mutation path, the rebinding step, and runner-side contract materialisation.

\subsection{Operational and Deployment Supplements}
\label{app:eval-ops}
The final supplementary evidence concerns behaviour that is operationally important but secondary to the decisive invariant tests in the main text. This material does not introduce new architectural claims. Instead, it shows how the same design behaves when viewed through recovery practice, decentralised claim, user-facing workflow compression, and deployment realism.

\subsubsection{Recovery Semantics and Administrative Requeue}

The experiment also sharpens the meaning of Invariant~D3. Exclusive ownership is often described as a concurrency property, but here it appears as a recovery property. If ownership were implicitly dissolved across restarts, then recovery would compete with the latent prior execution, and duplicate terminal behaviour would become possible. The absence of double execution after explicit requeue shows that ownership is not tied to the liveness of a particular agent process; it survives failure as part of the task's authoritative control-plane state. In this sense, exclusivity is not only about who may claim the work initially, but also about whether failure can fork the execution history. The present result shows that it cannot.

Finally, the experiment clarifies the philosophy of D4. ``Bounded recovery'' does not mean that the system autonomously repairs every failure as soon as possible. It means that once a failure is operationally recognised, there exists a well-typed, explicit, and idempotent path back into the normal lifecycle. That is exactly what the data show: one requeue per task, no duplicate execution, and subsequent completion through the standard claim path. The architecture therefore prefers \emph{correct recovery semantics} over automatic recovery theatre, which is the right trade-off for an HPC-facing service whose failures cross process, scheduler, and administrative boundaries.

\subsubsection{Multi-Agent Interleaving and Capacity Notes}

A second useful pattern is for the two runners to remain active throughout the burst, rather than partitioning the queue into front and back halves. Tasks assigned to \texttt{setonix-runner-1} span indices 0--49, while tasks assigned to \texttt{ella-runner-1} span indices 2--47. The claim process, therefore, remains interleaved across the entire run, which is consistent with decentralised competition at the claim boundary rather than one runner temporarily monopolising the queue.

The result is also more subtle than a throughput statement. If the protocol were semantically weak, the system might appear fast but risk duplicate claims or unstable ownership. If it were semantically correct but operationally clumsy, it might preserve exclusivity while underutilising the four available slots. The present data show neither pathology. Instead, \sys preserves the exact property that matters for this architecture: execution authority can be physically distributed across multiple runners and multiple scheduler submission paths without fragmenting task identity or collapsing control-plane efficiency.

Finally, this experiment changes the status of multi-agent concurrency in the overall evaluation. It is no longer merely a natural future extension of the architecture; it is now direct evidence that the exclusive-claim semantics developed for the single-agent case do not rely on single-agent centrality. The paper's concurrency story, therefore, closes at the right abstraction level: ownership remains authoritative even when execution is decentralised.

\subsubsection{Operational Comparison: \sys vs.\ Manual Slurm}
\label{sec:eval-manual}

The preceding experiments validate the semantic correctness of the service-export layer. The remaining question is what changes in that layer for the user boundary. The relevant comparison is not whether \sys is faster than direct scheduler access for an operator already inside the HPC environment, but whether it removes HPC-specific operational burden from the user path while preserving the same underlying simulation semantics and compute substrate.

\paragraph*{Protocol.}
We prepare 10 matched circuits (5 qubits, depth 5, 1024 shots, identical seeds, \texttt{ibm-fez-0309}) and execute them through two workflows.

\begin{itemize}[leftmargin=*,nosep]
  \item \textbf{\sys (automated):} a Python client submits each circuit through \texttt{POST /tasks} and polls for completion.
  \item \textbf{Manual Slurm:} an operator on the HPC login node prepares per-task directories, writes a standalone simulation script, generates \texttt{sbatch} files, submits via \texttt{sbatch}, polls \texttt{squeue}/\texttt{sacct}, and parses result files.
\end{itemize}

Both workflows use the same device snapshot, the same GPU hardware, and the same Qiskit-Aer backend. The comparison is therefore operational rather than algorithmic: the compute kernel is held fixed while the access boundary is changed.

\paragraph*{Results.}

\begin{table}[t]
  \centering
  \caption{Operational comparison between \sys and direct manual Slurm submission for 10 matched circuits.}
  \label{tab:manual-compare}
  \small
  \begin{tabular}{lrr}
    \toprule
    Metric & \sys & Manual Slurm \\
    \midrule
    Circuits completed               & 10/10 & 10/10 \\
    Mean $t_\text{simulate}$ (s)     & 0.0848 & 0.0798 \\
    Mean $t_\text{total-sim}$ (s)    & 1.9658 & 1.8524 \\
    Mean $T_\text{total}$ (s)        & 57.2 & 10.12 \\
    Batch wall time (s)              & 582.3 & 101.2 \\
    User-facing steps                & 2 & 9 \\
    HPC login               & No & Yes \\
    Module load             & No & Yes \\
    File management & No & Yes \\
    \bottomrule
  \end{tabular}
\end{table}

Both workflows complete all 10 circuits successfully. The simulation kernel itself is materially unchanged: mean $t_\text{simulate}$ remains in the 0.08--0.09\,s range in both paths, and the broader in-job execution time is also of the same order. Manual Slurm yields mean $t_\text{total-sim}=1.8524$\,s, while the automated path yields mean $t_\text{total-sim}=1.9658$\,s; this difference is largely explained by a one-time warm-up outlier on the first automated trial (\mbox{$t_\text{total-sim}=3.0965$\,s}), after which the remaining nine automated trials stabilise around 1.84\,s. The compute substrate is therefore effectively the same in both workflows.

The dominant difference lies in orchestration rather than computation. In the automated path, the mean end-to-end time is 57.2\,s, of which 56.8\,s is waiting. In the manual path, the mean end-to-end time is 10.12\,s, with 10.03\,s also spent waiting. In both cases, user-visible latency is governed overwhelmingly by observation and dispatch structure, not by the simulation kernel itself. What changes is where that orchestration resides.

The workflow boundary contracts sharply. Manual Slurm requires 9 explicit steps, including logging in, environment setup, per-task script preparation, batch script generation, submission, queue monitoring, and result parsing. The \sys path reduces this to 2 programmatic actions, a submission call and a status query, while eliminating all requirements for HPC login, module management, and per-task file handling.

\paragraph*{Interpretation.}
This experiment should not be read as a speed comparison in the narrow sense. For small workloads and under the current polling-based implementation, \sys is not faster than using the scheduler directly by an operator already within the HPC site. The more important result is architectural: the additional latency is in the orchestration layer, while the removed complexity is at the user boundary. \sys therefore trades direct scheduler immediacy for operational decoupling.

That trade-off is exactly what a service-export architecture is meant to accomplish. In the manual workflow, the user must personally cross the HPC boundary: authenticate to the site, understand the module environment, materialise scripts, manage files, and interpret scheduler state. In the \sys workflow, those responsibilities are absorbed by the service layer and realised by the agent within site policy. The user interacts with a stable device-oriented contract rather than with site-specific operational procedures. The experiment therefore validates Invariant~D1 in a practical sense: outbound-only export is not merely deployable, but meaningfully compresses the user path.

A deeper implication is that the comparison is not principally about seconds; it is about \emph{where expertise must reside}. The manual path assumes that every user is also an HPC operator. The \sys path encapsulates HPC-specific knowledge within a reusable systems layer. This is the practical meaning of service export in the present setting: not that it removes scheduler latency, but that it relocates scheduler-facing complexity from every client session to a controlled boundary component.

The timing structure is also consistent with the earlier latency analysis. The compute kernel remains effectively unchanged, while the additional delay in the automated path is attributable to deployment-level dispatch and polling behaviour rather than to altered simulation semantics. The manual comparison, therefore, does not weaken the earlier overhead argument; it sharpens it. \sys adds bounded orchestration latency in exchange for removing direct HPC dependency from the client access path.

\subsubsection{Deployment Feasibility}
\label{sec:eval-deployment}
A recurring objection to service-export architectures is that they often rely on operational assumptions that do not survive contact with real HPC deployments. The present evaluation answers that objection directly. Every experiment in this appendix was conducted on the production Pawsey deployment without privileged ingress changes, firewall exceptions, or administrator-managed inbound services. Submission, polling, event retrieval, device mutation, and recovery all traverse the same outbound-only coordination path assumed by Invariant~D1.

What matters here is not only that the system runs, but how it runs. The deployment results show that the architecture preserves a deliberate asymmetry between control and execution. The control plane may admit work, bind snapshots, and emit lifecycle events, but once a task has been claimed, execution proceeds from a local payload and bound snapshot rather than from live server consultation. This property is already implicit in the earlier experiments: the latency study decomposes runner-side execution from artifacts generated at the compute boundary; the snapshot-binding study depends on the claim-time contract remaining fixed once bound; and the crash-recovery study shows that post-claim semantics remain meaningful even when the agent disappears. Hermetic execution is therefore not merely asserted; it is exercised as a hidden prerequisite of the entire evaluation.

This gives Invariant~D8 a broader significance. Hermeticity is not only an implementation convenience for robustness; it is what makes the control-plane semantics trustworthy. If running jobs could consult mutable backend state during execution, then claim-time binding would cease to be a genuine commitment, recovery would become harder to reason about, and post hoc observability would no longer correspond cleanly to a single execution contract. By keeping the compute path closed over its local payload, the architecture prevents subsequent control-plane mutations from retroactively changing the semantics of an already-bound task.

The deployment results also support a portability claim, but in the correct sense. The site-specific elements in the current implementation are confined to the scheduler adapter, including the Slurm batch template and module environment. By contrast, the device model, lifecycle state machine, event semantics, administrative mutation path, and binding logic remain site-independent. The value of the Pawsey deployment is therefore not just that the prototype runs in a real machine room; it is that the experiment reveals where HPC specificity ends. The architecture localises site dependence to the operational boundary where it belongs, while preserving platform-independent semantics above that layer.

\subsection{Evaluation-to-Invariant Synthesis}
\label{app:eval-summary}

The preceding evaluation supplements provide numerical detail, control experiments, and operational context. Table~\ref{tab:eval-summary} gathers those strands back into one architecture-level view by linking each experimentally tested claim to the invariants it supports and to the implication that follows from the evidence.

\begin{table*}[t]
\centering
\caption{Evaluation synthesis. Each row links an experimentally tested claim to the design invariants from Section~\ref{main-sec:design} and to the architectural implication supported by the evidence.}
\vspace{-0.8em}
\label{tab:eval-summary}
\footnotesize
\setlength{\tabcolsep}{4pt}
\renewcommand{\arraystretch}{1.08}
\begin{tabularx}{\textwidth}{@{}p{0.18\textwidth}p{0.08\textwidth}Y Y@{}}
\toprule
Claim under test & Inv. & Operational evidence & Architectural implication \\
\midrule
Bounded service-path overhead &
D1, D4 &
$T_\text{admit}\!\approx\!0.4$\,s remains flat; interface overhead stays in the low-single-second range ($\approx$2--3.4\,s); workload growth appears only in the simulator term &
Service export adds latency but not a second scaling law: control-path cost remains bounded while workload scaling stays localised to the compute substrate \\

Calibration-aware execution &
D5, D6 &
Under identical topology and native gate set, TV changes from 0 (ideal snapshot) to 0.03537 (real Fez calibration), with low variance across 5 trials &
Snapshot content is part of the executable contract, not descriptive metadata; graph-parametric device interpretation remains valid on a 156-qubit heavy-hex device \\

Claim-time snapshot freshness &
D5, D7 &
In the same-device pre-claim mutation, 8/8 tasks execute with the post-mutation ideal snapshot rather than the submit-time noisy snapshot &
Freshness is resolved at the correct lifecycle boundary: queue delay does not freeze stale calibration semantics at submission \\

Exclusive claim under multi-agent concurrency &
D3 &
50/50 tasks complete under two concurrent agents; 24:26 split across runners; 7.45 tasks/min; batch wall time is within 3.3\% of the 4-slot idealised floor; no duplicate claim &
Atomic claim scales from local mutual exclusion to decentralised arbitration: multiple runners can approach slot-limited utilisation without a central dispatcher while preserving unique execution lineage \\

Authoritative lifecycle under failure &
D2, D3, D4 &
After agent crash, 3/3 tasks remain in \Running\ until explicit requeue; 3/3 later complete; one requeue per task; no double execution &
The DFA remains authoritative under failure, and recovery is explicit, auditable, and idempotent rather than emergent from heartbeat loss \\

Operational compression of HPC access &
D1 &
2 vs.\ 9 user-facing steps; no direct HPC login, module management, or per-task file handling; compute-side timing remains materially unchanged across paths &
The service layer relocates HPC-specific operational knowledge from each client session into a reusable boundary component, turning scheduler-native usage into device-oriented service access \\

Structural regression coverage &
D1--D9 &
258 deterministic tests across lifecycle, protocol, cache invalidation, SSE behaviour, and topology-parametric validation &
The deployment experiments rest on a structural correctness layer rather than on operational anecdotes alone \\
\bottomrule
\end{tabularx}
\end{table*}

In summary, the evaluation provides three layers of support for the architecture. First, it establishes \emph{semantic fidelity}: device snapshots function as executable contracts whose meaning is fixed at the correct lifecycle boundary. Second, it establishes \emph{control-plane authority}: ownership, lifecycle state, and recovery remain well-typed in the presence of concurrency and failure. Third, it establishes \emph{deployment realism}: these properties hold on production HPC infrastructure under outbound-only constraints, with the user path compressed into a service contract rather than exposed as raw scheduler practice.

\clearpage
\balance
\bibliographystyle{IEEEtran}
\bibliography{references}

\begin{thebibliography}{10}
\providecommand{\url}[1]{#1}
\csname url@samestyle\endcsname
\providecommand{\newblock}{\relax}
\providecommand{\bibinfo}[2]{#2}
\providecommand{\BIBentrySTDinterwordspacing}{\spaceskip=0pt\relax}
\providecommand{\BIBentryALTinterwordstretchfactor}{4}
\providecommand{\BIBentryALTinterwordspacing}{\spaceskip=\fontdimen2\font plus
\BIBentryALTinterwordstretchfactor\fontdimen3\font minus
  \fontdimen4\font\relax}
\providecommand{\BIBforeignlanguage}[2]{{%
\expandafter\ifx\csname l@#1\endcsname\relax
\typeout{** WARNING: IEEEtran.bst: No hyphenation pattern has been}%
\typeout{** loaded for the language `#1'. Using the pattern for}%
\typeout{** the default language instead.}%
\else
\language=\csname l@#1\endcsname
\fi
#2}}
\providecommand{\BIBdecl}{\relax}
\BIBdecl

\bibitem{preskill2018quantum}
J.~Preskill, ``Quantum computing in the nisq era and beyond,'' \emph{Quantum},
  vol.~2, p.~79, 2018.

\bibitem{Paraskevopoulos2025}
N.~Paraskevopoulos, D.~Hamel, A.~Sarkar, C.~G. Almudever, and S.~Feld, ``Arta:
  automating design space exploration of spin-qubit architectures,''
  \emph{Quantum Information Processing}, vol.~24, no.~6, p. 184, 2025.

\bibitem{Haener2016}
T.~H{\"a}ner, D.~S. Steiger, M.~Smelyanskiy, and M.~Troyer, ``High performance
  emulation of quantum circuits,'' in \emph{SC'16: Proceedings of the
  International Conference for High Performance Computing, Networking, Storage
  and Analysis}.\hskip 1em plus 0.5em minus 0.4em\relax IEEE, 2016, pp.
  866--874.

\bibitem{IBMQuantum2024}
{IBM Quantum}, ``Focus on utility-scale computing: cloud simulators and lab are
  now retired,''
  \url{https://quantum.cloud.ibm.com/announcements/en/product-updates/2024-05-15-2024-sunset-final-lab-simulators},
  2024, product update announcement, accessed 2026-03-17.

\bibitem{AWS2026}
{Amazon Web Services}, ``Amazon braket faqs,''
  \url{https://aws.amazon.com/braket/faqs/}, 2026, accessed: 2026-03-18. FAQ
  entry for the SV1 simulator.

\bibitem{babuji2019parsl}
Y.~Babuji, A.~Woodard, Z.~Li, D.~S. Katz, B.~Clifford, R.~Kumar, L.~Lacinski,
  R.~Chard, J.~M. Wozniak, I.~Foster \emph{et~al.}, ``Parsl: Pervasive parallel
  programming in python,'' in \emph{Proceedings of the 28th International
  Symposium on High-Performance Parallel and Distributed Computing}, 2019, pp.
  25--36.

\bibitem{rocklin2015dask}
M.~Rocklin \emph{et~al.}, ``Dask: Parallel computation with blocked algorithms
  and task scheduling.'' in \emph{SciPy}, 2015, pp. 126--132.

\bibitem{deelman2015pegasus}
E.~Deelman, K.~Vahi, G.~Juve, M.~Rynge, S.~Callaghan, P.~J. Maechling,
  R.~Mayani, W.~Chen, R.~F. Da~Silva, M.~Livny \emph{et~al.}, ``Pegasus, a
  workflow management system for science automation,'' \emph{Future Generation
  Computer Systems}, vol.~46, pp. 17--35, 2015.

\bibitem{wilde2011swift}
M.~Wilde, M.~Hategan, J.~M. Wozniak, B.~Clifford, D.~S. Katz, and I.~Foster,
  ``Swift: A language for distributed parallel scripting,'' \emph{Parallel
  Computing}, vol.~37, no.~9, pp. 633--652, 2011.

\bibitem{rajaraman2023frontier}
V.~Rajaraman, ``Frontier—world’s first exaflops supercomputer,''
  \emph{Resonance}, vol.~28, no.~4, pp. 567--576, 2023.

\bibitem{allen2025aurora}
\BIBentryALTinterwordspacing
W.~E. Allcock \emph{et~al.}, ``Aurora: Architecting argonne's first exascale
  supercomputer for accelerated scientific discovery,'' 2025. [Online].
  Available: \url{https://arxiv.org/abs/2509.08207}
\BIBentrySTDinterwordspacing

\bibitem{bayraktar2023cuquantum}
H.~Bayraktar, A.~Charara, D.~Clark, S.~Cohen, T.~Costa, Y.-L.~L. Fang, Y.~Gao,
  J.~Guan, J.~Gunnels, A.~Haidar \emph{et~al.}, ``cuquantum sdk: A
  high-performance library for accelerating quantum science,'' in \emph{2023
  IEEE International Conference on Quantum Computing and Engineering (QCE)},
  vol.~1.\hskip 1em plus 0.5em minus 0.4em\relax IEEE, 2023, pp. 1050--1061.

\bibitem{steiger2018projectq}
D.~S. Steiger, T.~H{\"a}ner, and M.~Troyer, ``Projectq: an open source software
  framework for quantum computing,'' \emph{Quantum}, vol.~2, p.~49, 2018.

\bibitem{ibm_quantum_learning}
\BIBentryALTinterwordspacing
{IBM Quantum}, ``Ibm quantum learning platform,'' IBM, Tech. Rep., 2026, online
  educational resource. [Online]. Available:
  \url{https://quantum.cloud.ibm.com/learning/en}
\BIBentrySTDinterwordspacing

\bibitem{ARCHER2Service2026}
{ARCHER2 Service}, ``Archer2 user guide: Data management and transfer,''
  \url{https://docs.archer2.ac.uk/user-guide/data/}, 2026, accessed:
  2026-03-18.

\bibitem{NERSCC2026}
{National Energy Research Scientific Computing Center}, ``Getting started at
  nersc,'' \url{https://docs.nersc.gov/getting-started/}, 2026, accessed:
  2026-03-18.

\bibitem{ORLCF2026}
{Oak Ridge Leadership Computing Facility}, ``Olcf policy guide,''
  \url{https://www.olcf.ornl.gov/for-users/olcf-policy-guide/}, 2026, accessed:
  2026-03-18.

\bibitem{PSRC2026}
{Pawsey Supercomputing Research Centre}, ``Getting started with
  supercomputing,''
  \url{https://pawsey.atlassian.net/wiki/spaces/US/pages/51925850/Getting+Started+with+Supercomputing},
  2026, accessed: 2026-03-18.

\bibitem{thain2005distributed}
D.~Thain, T.~Tannenbaum, and M.~Livny, ``Distributed computing in practice: the
  condor experience,'' \emph{Concurrency and computation: practice and
  experience}, vol.~17, no. 2-4, pp. 323--356, 2005.

\bibitem{turilli2018comprehensive}
M.~Turilli, M.~Santcroos, and S.~Jha, ``A comprehensive perspective on
  pilot-job systems,'' \emph{ACM Computing Surveys (CSUR)}, vol.~51, no.~2, pp.
  1--32, 2018.

\bibitem{Zheng2026}
\BIBentryALTinterwordspacing
D.~Zheng, J.~Xv, X.~Zhou, and Z.~Shan, ``Virtual qpu: A novel implementation of
  quantum computing,'' \emph{Computers, Materials and Continua}, vol.~87,
  no.~1, 2026. [Online]. Available:
  \url{https://www.sciencedirect.com/science/article/pii/S1546221826001712}
\BIBentrySTDinterwordspacing

\bibitem{yoo2003slurm}
A.~B. Yoo, M.~A. Jette, and M.~Grondona, ``Slurm: Simple linux utility for
  resource management,'' in \emph{Workshop on job scheduling strategies for
  parallel processing}.\hskip 1em plus 0.5em minus 0.4em\relax Springer, 2003,
  pp. 44--60.

\bibitem{SchedMD2026}
{SchedMD}, ``Slurm workload manager documentation,''
  \url{https://slurm.schedmd.com/documentation.html}, 2026, accessed:
  2026-03-18.

\bibitem{OpenPBSProject2026}
{OpenPBS Project}, ``Openpbs: Open source workload manager and job scheduler,''
  \url{https://www.openpbs.org/}, 2026, accessed: 2026-03-18.

\bibitem{QAD2026}
{Qiskit Aer Developers}, ``Qiskit aer documentation,''
  \url{https://qiskit.github.io/qiskit-aer/}, 2026, accessed: 2026-03-18.

\bibitem{JavadiAbhari2024}
A.~Javadi-Abhari, M.~Treinish, K.~Krsulich, C.~J. Wood, J.~Lishman, J.~Gacon,
  S.~Martiel, P.~D. Nation, L.~S. Bishop, A.~W. Cross \emph{et~al.}, ``Quantum
  computing with qiskit,'' \emph{arXiv preprint arXiv:2405.08810}, 2024.

\bibitem{Li2021}
A.~Li, B.~Fang, C.~Granade, G.~Prawiroatmodjo, B.~Heim, M.~Roetteler, and
  S.~Krishnamoorthy, ``Sv-sim: scalable pgas-based state vector simulation of
  quantum circuits,'' in \emph{Proceedings of the International Conference for
  High Performance Computing, Networking, Storage and Analysis}, 2021, pp.
  1--14.

\bibitem{Lykov2021}
D.~Lykov, A.~Chen, H.~Chen, K.~Keipert, Z.~Zhang, T.~Gibbs, and Y.~Alexeev,
  ``Performance evaluation and acceleration of the qtensor quantum circuit
  simulator on gpus,'' in \emph{2021 IEEE/ACM Second International Workshop on
  Quantum Computing Software (QCS)}.\hskip 1em plus 0.5em minus 0.4em\relax
  IEEE, 2021, pp. 27--34.

\bibitem{Li2019}
G.~Li, Y.~Ding, and Y.~Xie, ``Tackling the qubit mapping problem for nisq-era
  quantum devices,'' in \emph{Proceedings of the twenty-fourth international
  conference on architectural support for programming languages and operating
  systems}, 2019, pp. 1001--1014.

\bibitem{Murali2019}
P.~Murali, J.~M. Baker, A.~Javadi-Abhari, F.~T. Chong, and M.~Martonosi,
  ``Noise-adaptive compiler mappings for noisy intermediate-scale quantum
  computers,'' in \emph{Proceedings of the twenty-fourth international
  conference on architectural support for programming languages and operating
  systems}, 2019, pp. 1015--1029.

\bibitem{Kurniawan2024}
\BIBentryALTinterwordspacing
H.~Kurniawan, L.~Rodríguez-Soriano, D.~Cuomo, C.~G. Almudever, and F.~G.
  Herrero, ``On the use of calibration data in error-aware compilation
  techniques for nisq devices,'' 2024. [Online]. Available:
  \url{https://arxiv.org/abs/2407.21462}
\BIBentrySTDinterwordspacing

\bibitem{Haener2017}
T.~H{\"a}ner and D.~S. Steiger, ``5 petabyte simulation of a 45-qubit quantum
  circuit,'' in \emph{Proceedings of the International Conference for High
  Performance Computing, Networking, Storage and Analysis}, 2017, pp. 1--10.

\bibitem{Cai2023}
Z.~Cai, R.~Babbush, S.~C. Benjamin, S.~Endo, W.~J. Huggins, Y.~Li, J.~R.
  McClean, and T.~E. O’Brien, ``Quantum error mitigation,'' \emph{Reviews of
  Modern Physics}, vol.~95, no.~4, p. 045005, 2023.

\bibitem{Quek2024}
Y.~Quek, D.~Stilck~Fran{\c{c}}a, S.~Khatri, J.~J. Meyer, and J.~Eisert,
  ``Exponentially tighter bounds on limitations of quantum error mitigation,''
  \emph{Nature Physics}, vol.~20, no.~10, pp. 1648--1658, 2024.

\bibitem{Zhang2025}
A.~Zhang, H.~Xie, Y.~Gao, J.-N. Yang, Z.~Bao, Z.~Zhu, J.~Chen, N.~Wang,
  C.~Zhang, J.~Zhong \emph{et~al.}, ``Demonstrating quantum error mitigation on
  logical qubits,'' \emph{Nature Communications}, 2025.

\bibitem{AWS2025}
\BIBentryALTinterwordspacing
{Amazon Web Services}, ``{Amazon Braket: Quantum Computing Service},''
  \url{https://aws.amazon.com/braket/}, 2025, accessed: 2025-12-15. [Online].
  Available: \url{https://aws.amazon.com/braket/}
\BIBentrySTDinterwordspacing

\bibitem{IBMQuantum2026a}
{IBM Quantum}, ``Manage cost,''
  \url{https://quantum.cloud.ibm.com/docs/en/guides/manage-cost}, 2026,
  accessed: 2026-03-18.

\bibitem{Ma2025}
\BIBentryALTinterwordspacing
N.~Ma, H.~Li, N.~Ma, and H.~Li, ``Understanding and estimating the execution
  time of quantum circuits,'' New York, NY, USA, Nov. 2025, just Accepted.
  [Online]. Available: \url{https://doi.org/10.1145/3778031}
\BIBentrySTDinterwordspacing

\bibitem{PSRC2023}
{Pawsey Supercomputing Research Centre}, ``Setonix supercomputer,''
  \url{https://doi.org/10.48569/18sb-8s43}, Perth, Western Australia, 2023.

\bibitem{Armbrust2009}
\BIBentryALTinterwordspacing
M.~Armbrust, A.~Fox, R.~Griffith, A.~D. Joseph, R.~Katz, A.~Konwinski, G.~Lee,
  D.~Patterson, A.~Rabkin, I.~Stoica, and M.~Zaharia, ``A view of cloud
  computing,'' \emph{Commun. ACM}, vol.~53, no.~4, p. 50–58, Apr. 2010.
  [Online]. Available: \url{https://doi.org/10.1145/1721654.1721672}
\BIBentrySTDinterwordspacing

\bibitem{AWS2026a}
{Amazon Web Services}, ``Elastic fabric adapter,''
  \url{https://aws.amazon.com/hpc/efa/}, 2026, accessed: 2026-03-18.

\bibitem{AWS2026b}
------, ``Amazon fsx for lustre,'' \url{https://aws.amazon.com/fsx/lustre/},
  2026, accessed: 2026-03-18.

\bibitem{cockburn2005hexagonal}
A.~Cockburn, ``Hexagonal architecture,''
  \url{https://alistair.cockburn.us/hexagonal-architecture}, 2005, accessed:
  2026-03-17.

\bibitem{IBMQuantum2026}
{IBM Quantum}, ``Qiskit transpiler api documentation,''
  \url{https://quantum.cloud.ibm.com/docs/en/api/qiskit/transpiler}, 2026,
  accessed: 2026-03-18.

\bibitem{cirq_google}
{Google Quantum AI}, ``Cirq,'' \url{https://quantumai.google/cirq}, 2026,
  accessed: 2026-03-17.

\bibitem{Saurabh2023}
N.~Saurabh, S.~Jha, and A.~Luckow, ``A conceptual architecture for a
  quantum-hpc middleware,'' in \emph{2023 IEEE international conference on
  quantum software (QSW)}.\hskip 1em plus 0.5em minus 0.4em\relax IEEE, 2023,
  pp. 116--127.

\bibitem{Shehata2024}
A.~Shehata, T.~Naughton, and I.-S. Suh, ``A framework for integrating quantum
  simulation and high performance computing,'' in \emph{2024 IEEE International
  Conference on Quantum Computing and Engineering (QCE)}, vol.~2.\hskip 1em
  plus 0.5em minus 0.4em\relax IEEE, 2024, pp. 300--305.

\bibitem{Shehata2026}
A.~Shehata, P.~Groszkowski, T.~Naughton, M.~G. Meena, E.~Wong, D.~Claudino,
  R.~F. Da~Silva, and T.~Beck, ``Bridging paradigms: Designing for hpc-quantum
  convergence,'' \emph{Future Generation Computer Systems}, vol. 174, p.
  107980, 2026.

\bibitem{Mantha2025}
\BIBentryALTinterwordspacing
P.~Mantha, F.~J. Kiwit, N.~Saurabh, S.~Jha, and A.~Luckow, ``Pilot-quantum: A
  quantum-hpc middleware for resource, workload and task management,'' 2025.
  [Online]. Available: \url{https://arxiv.org/abs/2412.18519}
\BIBentrySTDinterwordspacing

\bibitem{pawsey_vqpu_hybrid_workflow}
{Pawsey Supercomputing Research Centre}, ``vqpu-hybrid-workflow: Hybrid (v)qpu,
  gpu, cpu workflow framework,''
  \url{https://github.com/PawseySC/vqpu-hybrid-workflow}, 2026, accessed:
  2026-03-18.

\bibitem{foster2006globus}
I.~Foster, ``Globus toolkit version 4: Software for service-oriented systems,''
  \emph{Journal of computer science and technology}, vol.~21, no.~4, pp.
  513--520, 2006.

\bibitem{galaxy2022galaxy}
{Galaxy Community}, ``The galaxy platform for accessible, reproducible and
  collaborative biomedical analyses: 2022 update,'' \emph{Nucleic acids
  research}, vol.~50, no.~W1, pp. W345--W351, 2022.

\bibitem{celery_docs}
{Celery Project}, ``Celery documentation: Distributed task queue,''
  \url{https://docs.celeryq.dev/en/stable/}, 2026, accessed: 2026-03-17.

\bibitem{rq_python}
{RQ Developers}, ``Rq: Simple job queues for python,''
  \url{https://python-rq.org/}, 2026, accessed: 2026-03-17.

\bibitem{tao2025quantum}
R.~Tao, H.~Zhu, J.~Nieh, J.~Yao, and R.~Gu, ``Quantum virtual machines,'' in
  \emph{19th USENIX Symposium on Operating Systems Design and Implementation
  (OSDI 25)}, 2025, pp. 411--428.

\bibitem{liu2026dynq}
S.~Liu, P.~J. Elahi, and U.~Varetto, ``Dynq: A dynamic topology-agnostic
  quantum virtual machine via quality-weighted community detection,''
  \emph{arXiv preprint arXiv:2601.19635}, 2026.

\bibitem{Lewis2023}
M.~Lewis, S.~Soudjani, and P.~Zuliani, ``Formal verification of quantum
  programs: Theory, tools, and challenges,'' \emph{ACM Transactions on Quantum
  Computing}, vol.~5, no.~1, pp. 1--35, 2023.

\bibitem{Wang2022}
H.~Wang, J.~Gu, Y.~Ding, Z.~Li, F.~T. Chong, D.~Z. Pan, and S.~Han,
  ``Quantumnat: quantum noise-aware training with noise injection, quantization
  and normalization,'' in \emph{Proceedings of the 59th ACM/IEEE design
  automation conference}, 2022, pp. 1--6.

\bibitem{LUMIConsortium2026}
{LUMI Consortium}, ``Lumi supercomputer,''
  \url{https://lumi-supercomputer.eu/}, 2026, accessed: 2026-03-18.

\end{thebibliography}

\end{document}